\documentclass[journal]{IEEEtran}
\usepackage{cite}
\ifCLASSINFOpdf
  \usepackage[pdftex]{graphicx}
\else
  \usepackage[dvips]{graphicx}
\fi
\usepackage{algorithm}
\usepackage[noend]{algpseudocode}
\usepackage{multirow}

\usepackage{amsmath}
\begin{document}
\title{Phoenix: A Low-Precision Floating-Point Quantization Oriented Architecture for Convolutional Neural Networks}

%\author{Michael~Shell,~\IEEEmembership{Member,~IEEE,}
%        John~Doe,~\IEEEmembership{Fellow,~OSA,}
%        and~Jane~Doe,~\IEEEmembership{Life~Fellow,~IEEE}% <-this % stops a space
%\thanks{M. Shell was with the Department
%of Electrical and Computer Engineering, Georgia Institute of Technology, Atlanta,

\author{Chen~Wu,
        Mingyu~Wang,
        Xiayu~Li,
        Jicheng~Lu,
        Kun~Wang,
        Lei~He
\thanks{C. Wu, M. Wang, K. Wang and L. He are with University of California, Los Angeles, 90095, CA, USA (e-mail: 
        chenwu1989@ucla.edu, mingyuw@ucla.edu, wangk@ucla.edu, lhe@ee.ucla.edu).}
\thanks{X. Li and J. Lu are with Shanghai Fudan Microelectronics Group Company Limited, Shanghai, China (e-mail:
        lixiayu@fmsh.com.cn, lujicheng@fmsh.com.cn).}}

\maketitle

\begin{abstract}
Convolutional neural networks (CNNs) achieve state-of-the-art performance at the cost of becoming deeper and larger. Although quantization (both fixed-point and floating-point) has proven effective for reducing storage and memory access, two challenges -- 1) accuracy loss caused by quantization without calibration, fine-tuning or re-training for deep CNNs and 2) hardware inefficiency caused by floating-point quantization -- prevent processors from completely leveraging the benefits.
In this paper, we propose a low-precision floating-point quantization oriented processor, named {\sf Phoenix}, to address the above challenges. We primarily have three key observations: 1) 8-bit floating-point quantization incurs less error than 8-bit fixed-point quantization; 2) without using any calibration, fine-tuning or re-training techniques, normalization before quantization further reduces accuracy degradation; 3) 8-bit floating-point multiplier achieves higher hardware efficiency than 8-bit fixed-point multiplier if the full-precision product is applied. Based on these key observations, we propose a normalization-oriented 8-bit floating-point quantization method to reduce storage and memory access with negligible accuracy loss (within 0.5\%/0.3\% for top-1/top-5 accuracy, respectively). We further design a hardware processor to address the hardware inefficiency caused by floating-point multiplier. Compared with a state-of-the-art accelerator, {\sf Phoenix} is 3.32$\times$ and 7.45$\times$ better in performance with the same core area for AlexNet and VGG16, respectively.
\end{abstract}


% \fi
\IEEEpeerreviewmaketitle

\section{Introduction}
\label{section:introduction}
\IEEEPARstart{C}{onvolutional} neural networks and deep neural networks have demonstrated a breakthrough in performance for a broad range of applications, including object recognition \cite{obj_recog}, object detection \cite{obj_detec} and speech recognition \cite{spe_recog}. The advantages mainly come from the huge computational complexity and huge amount of data. This motivates researchers in both academia and industry to focus on accelerating CNNs by using CPU/GPU clusters \cite{cluster}, FPGAs \cite{brainwave} and ASICs \cite{diannao}. Among them, customized accelerators/processors on FPGAs and ASICs have shown more promising throughput and energy efficiency than traditional CPU/GPU clusters \cite{caffeine, systolic, dadiannao, cambricon}.

Meanwhile, algorithm researchers keep on designing larger and deeper CNNs to improve performance in a broader range of scenarios. Such networks use larger amount of parameters, e.g. AlexNet \cite{alexnet} in 2012 has 249.51MB parameters, while VGG-16 \cite{vgg} in 2014 has 553.43MB parameters. The great quantities of parameters lead to a big challenge for communication between off-chip and on-chip memory because of bandwidth constraints. On the other hand, algorithm developers also focus on reducing parameters while maintaining performance by introducing residual \cite{resnet_v1} or inception blocks \cite{inception_v3, inception_v4}. They successfully decrease the parameter size dramatically, e.g., from 553.43MB in VGG16 to 68.63MB in DenseNet201 \cite{densenet}. However, the number of {\it convolutional} layers or {\it fully-connected} layers increases significantly from 13 to 201, making a deeper CNN. Such deep CNNs render approximate computing ({\it e.g.,} quantization) for CNNs more difficult, as we need to balance the approximation error imposed by more layers \cite{qual_fix}.

\begin{table*}[t]
  \centering
  \caption{Comparison of existing quantization methods and corresponding accelerators.}
  \begin{tabular}{|c|c|c|c|c|c|}  \hline
                               &{\bf Slim} &{\bf Medium} &{\bf Deep} &{\bf Hardware}         &{\bf Note}                                           \\ \hline \hline
    Nvidia \cite{nvidia_fix}   &$\surd$    &$\surd$      &$\surd$    &GPU                    &Fixed-point, extra cost for calibration                     \\ \hline
    OLAccel \cite{OLAccel}     &$\surd$    &$\surd$      &$\surd$    &Customized Accelerator &Fixed-point, extra cost for re-training                     \\ \hline
    ARM \cite{ARM_mixed}       &$\surd$    &$\surd$      &x          &CPU                    &Fixed-point and floating-point mixed, no hardware support   \\ \hline
    Xilinx \cite{xilinx_float} &$\surd$    &$\surd$      &x          &CPU                    &No hardware support, extra cost for calibration             \\ \hline
    BFP \cite{bfp_fpga}        &$\surd$    &$\surd$      &x          &Customized Processor   &Block floating-point                                        \\ \hline
    \textsf{Phoenix}           &$\surd$    &$\surd$      &$\surd$    &Customized Processor   &Floating-point, no calibration, fine-tuning nor re-training \\ \hline
  \end{tabular}
  \label{table:support_cmp}
\end{table*}

Progress has been made to alleviate bandwidth constraints by reducing the amount of parameters. Quantization, which approximates full precision parameters with low-precision numbers, has emerged as an efficient solution among various techniques. 8-bit fixed-point quantization is one of the most commonly used techniques \cite{roofline, tpu, diannao}, which results in a 4$\times$ data reduction. More aggressive studies tried to further reduce the data size by introducing binary neural network \cite{bin1} or ternary neural network \cite{ter1}. Meanwhile, the authors in \cite{xilinx_float, nvidia_fix, entropy_quan, V-Quant} focused on maintaining comparable accuracy for deeper CNNs during quantization, by using calibration, fine-tuning or re-training techniques with training data after quantization. 

However, quantization with calibration, fine-tuning or re-training requires extra computing and training data which could be difficult for hardware developers. Xilinx in \cite{xilinx_float} proposed a low-precision floating-point quantization method with a single inference batch for calibration. They only proved their effectiveness for {\it slim} CNNs, which is not enough because deeper CNNs, {\it e.g.,} ResNet152 \cite{resnet_v1} and DenseNet201 \cite{densenet}, are more popular to gain higher performance in different applications. Moreover, Xilinx did not present any hardware for their quantization method. This is because floating-point multiply-accumulators (MACs) have lower hardware efficiency than fixed-point ones. This is another challenge for accelerator design.

In this paper, we propose a cooperative software/hardware approach to overcome the above challenges efficiently. Initially, we observe that non-uniform quantization for both weights and activations (we use activations to represent the outputs of a layer) incurs less quantization error compared with uniform quantization. At the same time, applying normalization on activations before quantization can reduce accuracy loss without any calibration, fine-tuning or re-training processes. Moreover, 8-bit floating-point multiplier achieves higher hardware efficiency than 8-bit fixed-point one when maintaining full precision results after multiplication. Therefore, we propose a normalization-oriented 8-bit floating-point quantizer to reduce the data size with negligible accuracy loss. Our proposed quantizer works for deep CNNs (more than 100 {\it convolutional/fully-connected} layers). On average, the top-1 accuracy loss is within 0.5\%, while the state-of-the-art work \cite{V-Quant} which can reach to such {\it deep} CNNs has a top-1 accuracy loss about 1\% with fine-tuning. After reducing the memory and bandwidth requirements by using 8-bit floating-point quantization method, we design a floating-point based hardware processor, named {\sf Phoenix}, to further address the problem of hardware efficiency caused by floating-point MACs. We maintain full precision for intermediate results after 8-bit floating-point multiplier, which saves 8.14$\times$ area compared with the 8-bit fixed-point multiplier. Compared with a state-of-the-art accelerator \cite{eyeriss_jssc}, {\sf Phoenix} is 3.32$\times$ and 7.45$\times$ better for AlexNet and VGG16 in terms of performance when considering the same core area at TSMC 65nm technology, respectively. {\sf Phoenix} also achieves 151$\times$ better in terms of energy compared with Nvidia TITAN Xp GPU with single image inference.

Our main contributions can be summarized as follows: 1) We make the key observations that normalization and non-uniform quantization can help reduce the quantization error even in {\it deep} CNNs. 2) Based on the observations, we propose a normalization-oriented 8-bit floating-point quantization method that dramatically reduces the parameter size with negligible accuracy loss for {\it deep} CNNs. 3) We further design a floating-point oriented hardware processor, named {\sf Phoenix}, which is placed and routed in TSMC 28nm technology, to solve the hardware inefficiency caused by floating-point MACs.

\section{Background and Motivation}
\label{section:background and motivation}

\subsection{Background}
\label{subsection:background}

\subsubsection{CNNs}
\label{subsubsection:cnn}

CNNs are used to classify or recognize objects by passing the inputs through multiple types of layers. In each layer, multiple neurons are constructed to process different inputs and pass the outputs to the next layer through connections, which store the weights for the network. Based on different processing procedures, the layers are typically divided into {\it convolutional, pooling, activation, normalization, fully-connected, residual} and {\it inception} layers. Among them, {\it convolutional/fully-connected} layers consume most portions of computation while {\it fully-connected} layers require largest memory to store weights. According to this, we divide the size of CNNs into three categories with respect to the number of {\it convolutional/fully-connected} layers: {\it slim} for less than 50 layers, {\it medium} for 50 to 100 layers, and {\it deep} for more than 100 layers (as shown in Table~\ref{table:benchmark} where we report the detailed network information).

\subsubsection{Over Parameterization in Neural Networks}
\label{subsubsection:over_parameterization}

Modern CNNs show dominant performance advantages in various application domains by enlarging the network architecture. However, such technique results in a heavy burden for memory capacity, communication bandwidth, computation and communication energy. Existing work tried to overcome such challenges, including algorithm-level techniques ({\it e.g.,} dropout \cite{dropout}, low-precision training \cite{low_training, IBM}) and architecture-level techniques ({\it e.g.,} approximate computing \cite{approx_computing}, low-precision operation \cite{bin2, ter2}). Among them, quantization, one of the approximate computing techniques, turns out to be effective. 8-bit fixed-point quantization is widely used by FPGA or ASIC for {\it slim} and {\it medium} CNNs \cite{shidiannao, EIE, en_dsp, bit_fusion}. Recently, it has advanced to {\it deep} CNNs, {\it e.g.,} ResNet101 on GPU for negligible accuracy loss with calibration \cite{nvidia_fix}. 8-bit floating-point quantization still stays in the algorithm-level optimization and can apply only to {\it medium} CNNs. More aggressively, binary neural and ternary neural networks are proposed to further reduce memory, bandwidth and energy but at the cost of a larger accuracy loss \cite{xnor_net, FINN, tnn}.

\subsubsection{8-bit Floating-Point}
\label{subsubsection:8_fp}
Similar to the definition of 32-bit floating-point from the IEEE-754 standard \cite{IEEE754}, the binary representation of 8-bit floating-point number comprises {\it sign, mantissa} and {\it exponent} in order. The decimal value of 8-bit floating-point number is then calculated by:
\begin{equation}
\label{equation:8_float}
    V_{dec} = (-1)^S\times1.M\times2^{E-bias},
\end{equation}
where $V_{dec}$ is the value in decimal, $S, M$ and $E$ are all unsigned values and denote the {\it sign, mantissa} and {\it exponent}, respectively. For $bias$ in Eq. ~(\ref{equation:8_float}), it is introduced for both positive and negative exponents as
\begin{equation}
\label{equation:bias}
    bias = 2^{E_{w}-1}-1,
\end{equation}
where $E_w$ is the data width of $E$. The data widths for $M$ and $E$ are not fixed for 8-bit floating-point and we will emulate all combinations. In the later sections, we use the term $MaEb$ to indicate different combinations, where $a$ and $b$ indicate the bit width of $M$ and $E$, respectively, e.g., $M3E4$ means the mantissa is 3 bits while the exponent is 4 bits. 

There are three special definitions in IEEE-754 standard. The first is subnormal numbers when $E=0$, and Eq. ~(\ref{equation:8_float}) is modified to:
\begin{equation}
\label{equation:subnormal}
    V_{dec} = (-1)^S\times0.M\times2^{1-bias}.
\end{equation}
Then, Infinity (Inf) and Not a Number (NaN) are the other two special cases, but not used in our work. This is because our saturation scheme saturates large numbers to the maximal number as illustrated in detail in Section~\ref{subsection:quantization_process}.

\subsection{Motivation}
\label{subsection:motivation}

\begin{figure}[t]
\centering
\includegraphics[width=3.5in]{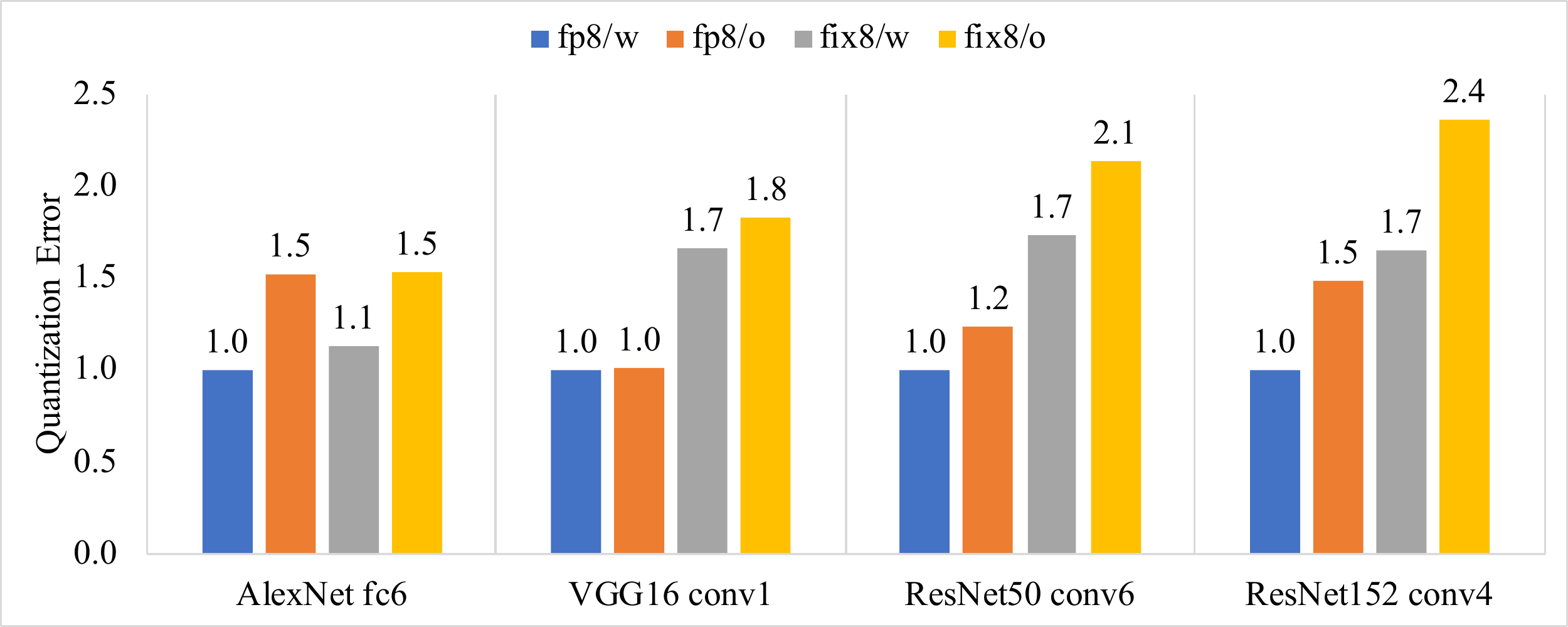}
\caption{Quantization error compared among floating-point and fixed-point quantizers with/without normalization (w: with normalization, o: without normalization).}
\label{fig:ob_quan_err}
\end{figure}

\subsubsection{Non-uniform quantizer}
\label{subsubsection:non-uniform}

Non-uniform quantizer has been proven more accurate than uniform quantizer when the inputs follow the non-uniform distributions, such as Gaussian, Laplacian, and Gamma distributions \cite{book_img, max_quan, paez_quan}. Although the weights and activations in a CNN are more likely to follow a non-uniform distribution, few efforts have been done to explore non-uniform quantizer for CNNs. Uniform quantizers, {\it e.g.,} fixed-point quantizers, are exploited instead on both CPU/GPU \cite{google_fix, nvidia_fix} and customized accelerators \cite{Stripes, bit_fusion, BPR} because of their hardware efficiency. A few researchers recently support non-uniform quantizer \cite{OLAccel, ARM_mixed, xilinx_float, bfp_fpga}. However, they either require notably extra cost to compensate for quantization error or fail to have efficient hardware, as shown in Table~\ref{table:support_cmp}. Particularly, the work in \cite{nvidia_fix} first reached {\it deep} CNNs, {\it e.g.,} ResNet101, with 8-bit fixed-point quantization by using calibration to compensate for quantization error. OLAccel \cite{OLAccel} divided weights and activations into the low-precision region (97\%) and high-precision region (3\%). The data in low-precision region were then quantized to 4 bits while the data in high-precision region remained full precision. Re-training is required in order to maintain accuracy, otherwise the proportion of high-precision region must be increased, which leads to higher cost on hardware implementation. ARM \cite{ARM_mixed} proposed a mixed quantizer, which used floating-point to represent weights and fixed-point to approximate activations. However, their approach remains in {\it medium} CNNs and lacks hardware support. Recently, Xilinx \cite{xilinx_float} developed the first low-precision floating-point quantization method. However, this approach needs to calibrate with an extra inference batch to maintain accuracy, and only {\it slim} and {\it medium} CNNs, {\it e.g.,} VGG16 and ResNet50, are validated. Moreover, the performance of running inference on a network is even worse than that of a full-precision network due to lack of hardware support. Block floating-point \cite{bfp_fpga} divided data into different data blocks, in which the data had different mantissas and a shared exponent. However, this approach was only validated for {\it medium} CNNs. To conclude, existing quantization algorithms and corresponding architectures cannot completely benefit from non-uniform quantizers.

\subsubsection{Observation}
\label{subsubsection:observation}

Previous quantization methods ignore the distribution properties of weights and activations. We fully analyze the distributions of weights and activations in different CNNs, and have two key observations: (1) A non-uniform quantizer fits weights and activations better than uniform quantizer as the weights and activations are more likely to follow Gaussian distributions; (2) Normalizing activations before quantization can further reduce quantization error. We select four representative layers -- {\it fc6} in AlexNet, {\it conv1} in VGG16, {\it conv6} in ResNet50 and {\it conv4} in ResNet152 as driving examples. The quantization errors caused by 8-bit floating-point and 8-bit fixed-point quantizers with/without normalization are depicted in Figure~\ref{fig:ob_quan_err}. We normalize the quantization error of all the cases with respect to the error caused by 8-bit floating-point quantizer with normalization. As can be seen from Figure~\ref{fig:ob_quan_err}, 8-bit floating-point quantizer with normalization incurs the lowest quantization error among all the four cases. Fixed-point quantizer with/without normalization causes 1.50$\times$/1.54$\times$ larger quantization error than 8-bit floating-point quantizer on average, respectively. In addition, with normalization, 8-bit floating-point and fixed-point quantizers both incur 1.3$\times$ less error. More detailed results about classification accuracy for CNNs will be illustrated in Subsection~\ref{subsection:normalization_analysis}. 

As for the hardware implementation, we evaluate the hardware efficiency of 8-bit floating-point and 8-bit fixed-point multipliers by implementing them with Verilog and synthesizing with Synopsys Design Compiler (DC). We get another key observation that if we keep full precision for the intermediate results after multiplication, 8-bit floating-point multiplier consumes less area than 8-bit fixed-point multiplier. For instance, 8-bit floating-point multiplier with $M4E3$ consumes only 12.3\% area of the 8-bit fixed-point multiplier. This is because $M4E3$ only needs a 5-bit unsigned fixed-point multiplier and a 3-bit unsigned fixed-point adder. The area savings are more significant for 8-bit floating-point multipliers with less mantissa bits (as shown in Figure~\ref{fig:mul}). In short, one can design a hardware efficient processor with 8-bit floating-point multiplier, and maintain accuracy at the same time by using 8-bit floating-point quantizer with normalization.

\section{Quantized Neural Network}
\label{section:qnn}

In this section, we present the details of our proposed normalization-based 8-bit floating-point quantization method, including quantization process, quantization results and normalization analysis.

\subsection{Quantization Process}
\label{subsection:quantization_process}

\begin{figure}[t]
\centering
\includegraphics[width=3.5in]{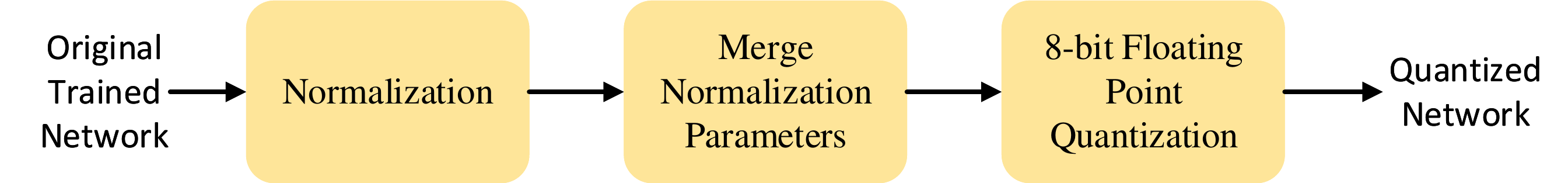}
\caption{Flow of 8-bit floating-point quantization.}
\label{fig:quan_process}
\end{figure}

Instead of quantizing the network directly, our quantization method is divided into three steps: normalization, merge normalization parameters, and quantization, as shown in Figure~\ref{fig:quan_process}. 

\subsubsection{Normalization}
\label{subsubsection:normalization}

As illustrated in Subsection~\ref{subsection:motivation}, normalizing activations can reduce quantization error. Therefore, we first normalize activations based on the inference results on the original network. The normalization method is defined as (we exploit other normalization methods in Subsection~\ref{subsection:normalization_analysis})
\begin{equation}
\label{equation:normlization}
    NORM\_O_i^m = \frac{O_i^m}{\sqrt{E((O^m)^2)}},
\end{equation}
where $O_i^m$ and $NORM\_O_i^m$ indicate the $i$th activation before and after normalization for layer $m$, respectively; $E((O^m)^2)$ is the second moment of the activations, expressed as 
\begin{equation}
\label{equation:2nd_moment}
    E((O^m)^2) = \frac{1}{N}\sum_{j=0}^{N - 1}(O_j^m)^2,
\end{equation}
where $N^m$ denotes the total number of activations of layer $m$ with $N^m=OH^m \times OW^m \times OC^m$, $OH^m, OW^m$ and $OC^m$ being the height, width and channel number of the output layer, respectively.

\subsubsection{Merge Normalization Parameters}
\label{subsubsection:merge}

\begin{figure}[t]
\centering
\includegraphics[width=3.5in]{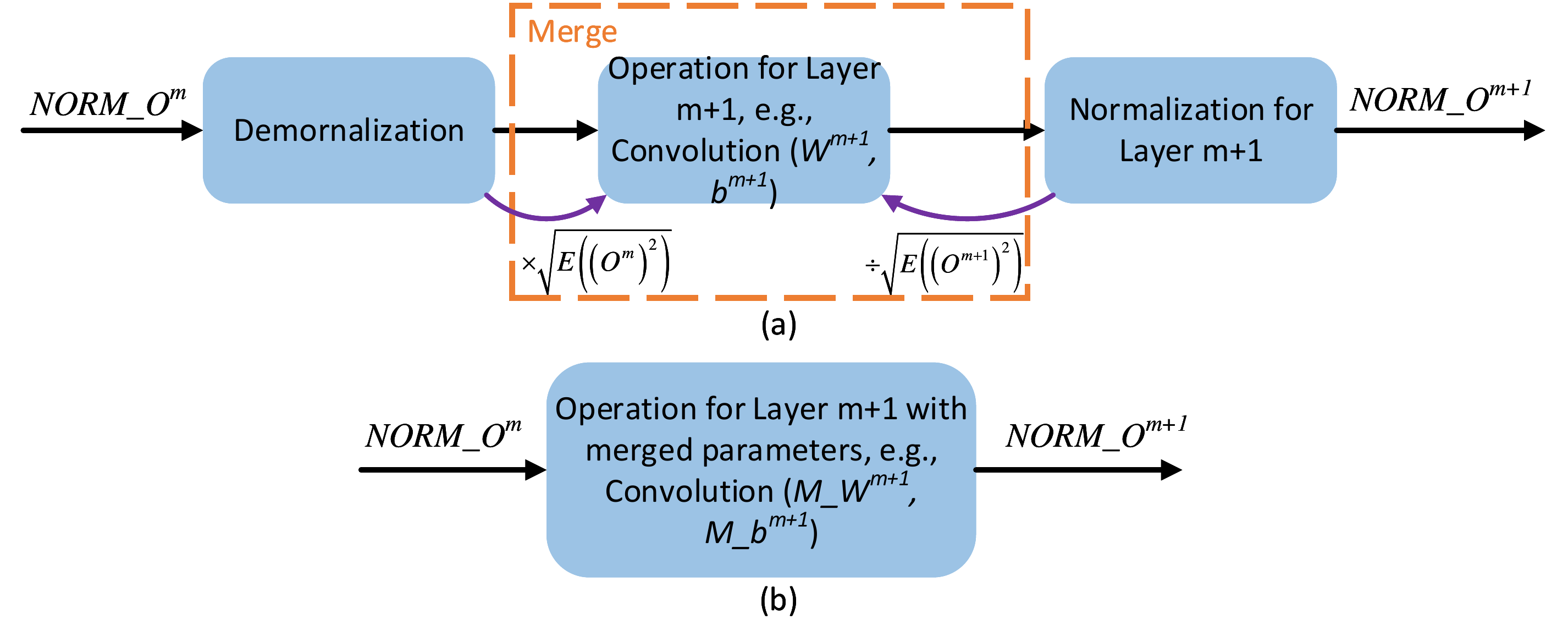}
\caption{(a) Flow of merging normalization parameters. (b) Simplified calculation process after merging all the normalization parameters.}
\label{fig:merge_norm}
\end{figure}

As normalization is utilized to reduce quantization error in our proposed approach, it does not incur any accuracy loss. Therefore, we merge all the normalization parameters into the parameters of each layer, as shown in Figure~\ref{fig:merge_norm}(a). As can be seen in Figure~\ref{fig:merge_norm}(a), denormalization is first applied to the normalized outputs of layer $m$ to make sure that normalization for layer $m$ never incurs accuracy loss. After the operation ( {\it e.g.,} {\it convolution, pooling, activation}) of layer $m+1$, the outputs need to be normalized in order to reduce quantization error. As both the denormalization and normalization procedures are linear, they can be merged to the parameters of layer $m+1$, as marked with dashed line in Figure~\ref{fig:merge_norm}(a). To clearly explain the merging procedure, we use {\it convolutional} and {\it fully-connected} layers as driving examples. In the $m$-th {\it convolutional} layer or {\it fully-connected} layer, output neurons are calculated as
\begin{equation}
\label{equation:output}
    O^m = W^m \cdot I^m + b^m,
\end{equation}
where $W^m, I^m$ and $b^m$ denote the weight matrix, input matrix and bias vector for layer $m$, respectively. Since the outputs of layer $m$ are normalized with Eq. ~(\ref{equation:normlization}) and fed as the inputs of layer $m+1$, the denormalization process can be applied to the weights of layer $m+1$ as follows:
\begin{equation}
\label{equation:denorm}
    DE\_W^{m+1} = W^{m+1} \times \sqrt{E((O^m)^2)},
\end{equation}
where $DE\_W^{m+1}$ means the denormalized weights of layer $m+1$. In order to simplify the calculation of normalization for layer $m+1$, we also merge the normalization into the weights of layer $m+1$, formulated as
\begin{subequations}
\label{equation:merge}
    \begin{align}
        M\_W^{m+1} = DE\_W^{m+1} / \sqrt{E((O^{m+1})^2)} \\
        M\_b^{m+1} = b / \sqrt{E((O^{m+1})^2)},
    \end{align}
\end{subequations}
where $M\_W^{m+1}$ and $M\_b^{m+1}$ are the weights and biases of layer $m+1$ with both denormalization and normalization. After merging denormalization and normalization into the weights and biases, we can get the normalized outputs with a simplified calculation process, as shown in Figure~\ref{fig:merge_norm}(b).

\subsubsection{Quantization}
\label{subsubsection:quantization}

8-bit float point quantization is then applied to the activations and weights of the normalized network to save storage, and it is illustrated as follows:
\begin{equation}
\label{equation:quantization}
    V_{fp8} = round(V_{fp32} \times 2^{h\_s}, -MAX_{fp8}, MAX_{fp8}),
\end{equation}
where $V_{fp8}$ and $V_{fp32}$ represent the 8-bit floating-point value and 32-bit floating-point value, respectively; $MAX_{fp8}$ indicates the maximal number which can be represented with 8-bit floating-point, and $h\_s$ is the scaling factor to fit the data into the dynamic range of 8-bit floating-point data representation. The $round$ function in Eq. ~(\ref{equation:quantization}) rounds the data to the nearest value with saturation considered, formulated as
\begin{equation}
\label{equation:round}
    round(x, MIN, MAX) = 
    \begin{cases}
        MIN &x <= MIN \\
        MAX &x >= MAX \\
        round(x) &\text{otherwise}
    \end{cases}
    ,
\end{equation}
where $MIN$ and $MAX$ are the minimal and maximal values, respectively. 

During the quantization process, our aim is to find the optimal scaling factor to minimize the mean square error (MSE) compared with the full precision results, as illustrated by the pseudo-code in Algorithm~\ref{algo:quantization}. In our proposed quantization method, both the weights and activations are quantized. Since all the activations are normalized, we can set the scaling factor $h\_s$ the same for all the layers, which also simplifies the architecture design, especially for {\it residual} and {\it inception} layers. In the {\it residual} layer, element-wise addition are performed to the outputs of two previous layers, while in the {\it inception} layer, the outputs of several previous layers are concatenated into a single layer. If there are more than two layers, the hidden scale should be the same for all layers. This simplifies the architecture design.

\subsubsection{Data flow in processor}
\label{subsubsection:dataflow}

\begin{figure}[t]
\centering
\includegraphics[width=3.5in]{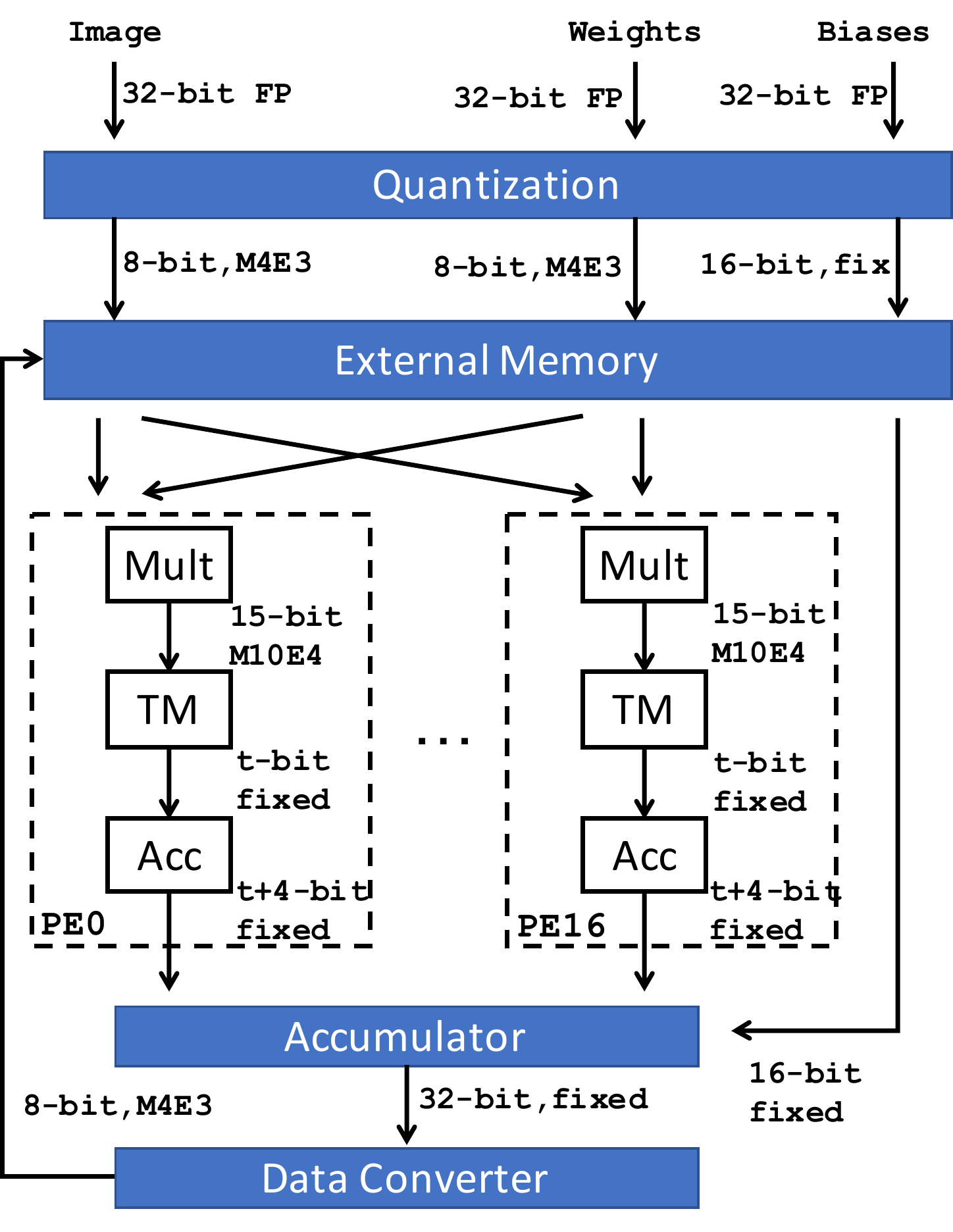}
\caption{The data flow in {\sf Phoenix} with $M4E3$ data format as an example (FP: floating-point, Mult: 8-bit floating-point multiplier, TM: truncating module, Acc: accumulator).}
\label{fig:data_flow}
\end{figure}

The data flow to run inference of a quantized network in {\sf Phoenix} is shown in Figure~\ref{fig:data_flow}. In order to explicitly illustrate the data flow, we list the bit width in each step with $M4E3$ data format as an example. All the input image, weights and biases are represented by 32-bit floating-point. In our processor, the original input image and weights are quantized with $M4E3$ data format and stored in external memory, while biases are quantized to 16-bit fixed-point to reduce quantization error. Multiplications are performed with the quantized image and weights, and the 15-bit floating-point ($M10E4$) products are aligned and truncated to $t$-bit fixed-point in the truncating module (the selection of $t$ will be explained in Subsubsection~\ref{subsubsection:truncating_bit}). In this way, all the accumulation can be done in fixed-point accumulators, which consumes fewer resources than floating-point accumulators. The final outputs in each output channel are converted to $M4E3$ floating-point again (and stored in the external memory) before used by another CNN layer. In the data flow, the truncating module and final data conversion step introduce bit truncation and precision loss. However, the precision loss has little impact on the final accuracy and is validated in Subsection~\ref{subsection:quantization_results} with comprehensive experimental results. 

\begin{algorithm}[t]
    \caption{Quantization}
    \label{algo:quantization}
    \begin{algorithmic}[1]
        \While{$m <= \# \text{ of layer}$}
            \State $i \gets -10$
            \While{$i < 10$}
                \State $W_{fp8} \gets round(W_{fp32}\times 2^{i}, -MAX_{fp8}, MAX_{fp8})$
                \State $MSE \gets \frac{1}{N}\sum_{k=0}^N (W_{fp8} - W_{fp32})^2$
                \If{$MSE < MSE_{min}$} 
                    \State $h\_s(m) \gets i$
                    \State $MSE_{min} \gets MSE$
                \EndIf
                \State $i \gets i+1$
            \EndWhile
            \State $m \gets m+1$
        \EndWhile
        \State \textbf{return} $h\_s$
    \end{algorithmic}
\end{algorithm}

\subsection{Quantization Results}
\label{subsection:quantization_results}

We implement our quantization method with C language based on the Darknet framework \cite{darknet}, by which the validation accuracy with single center-crop is evaluated with the ImageNet validation set (50,000 labelled images) \cite{imagenet}. Our quantization process is run on an Intel (R) Core (TM) i9-7960X CPU working under 2.86GHz, while the evaluation process is run on a Nvidia TITAN Xp GPU.

\begin{figure}[t]
\centering
\includegraphics[width=3.5in]{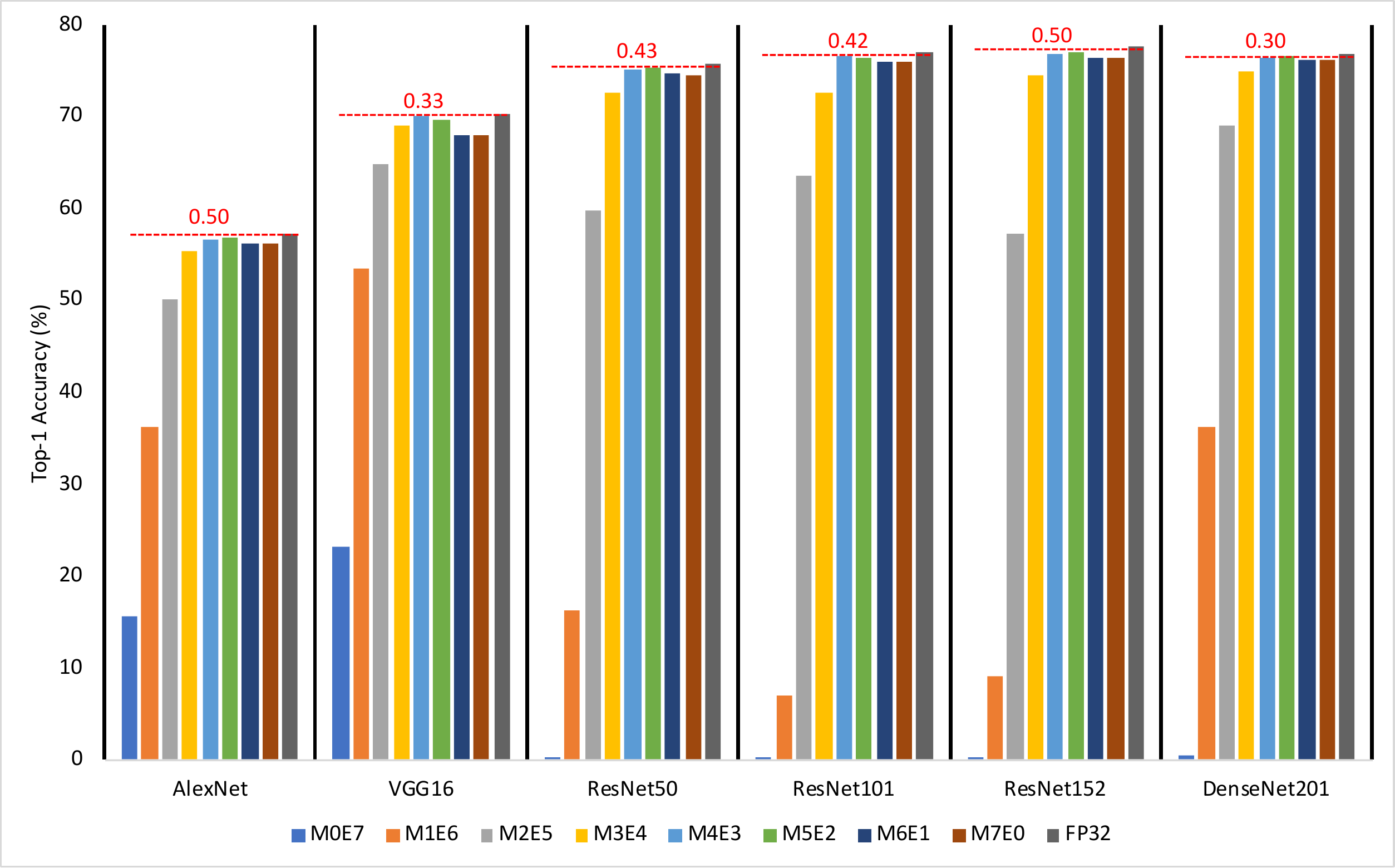}
\caption{Top-1 accuracy for different (mantissa, exponent) combinations with respect to different CNNs.}
\label{fig:top_1}
\end{figure}

\begin{figure}[t]
\centering
\includegraphics[width=3.5in]{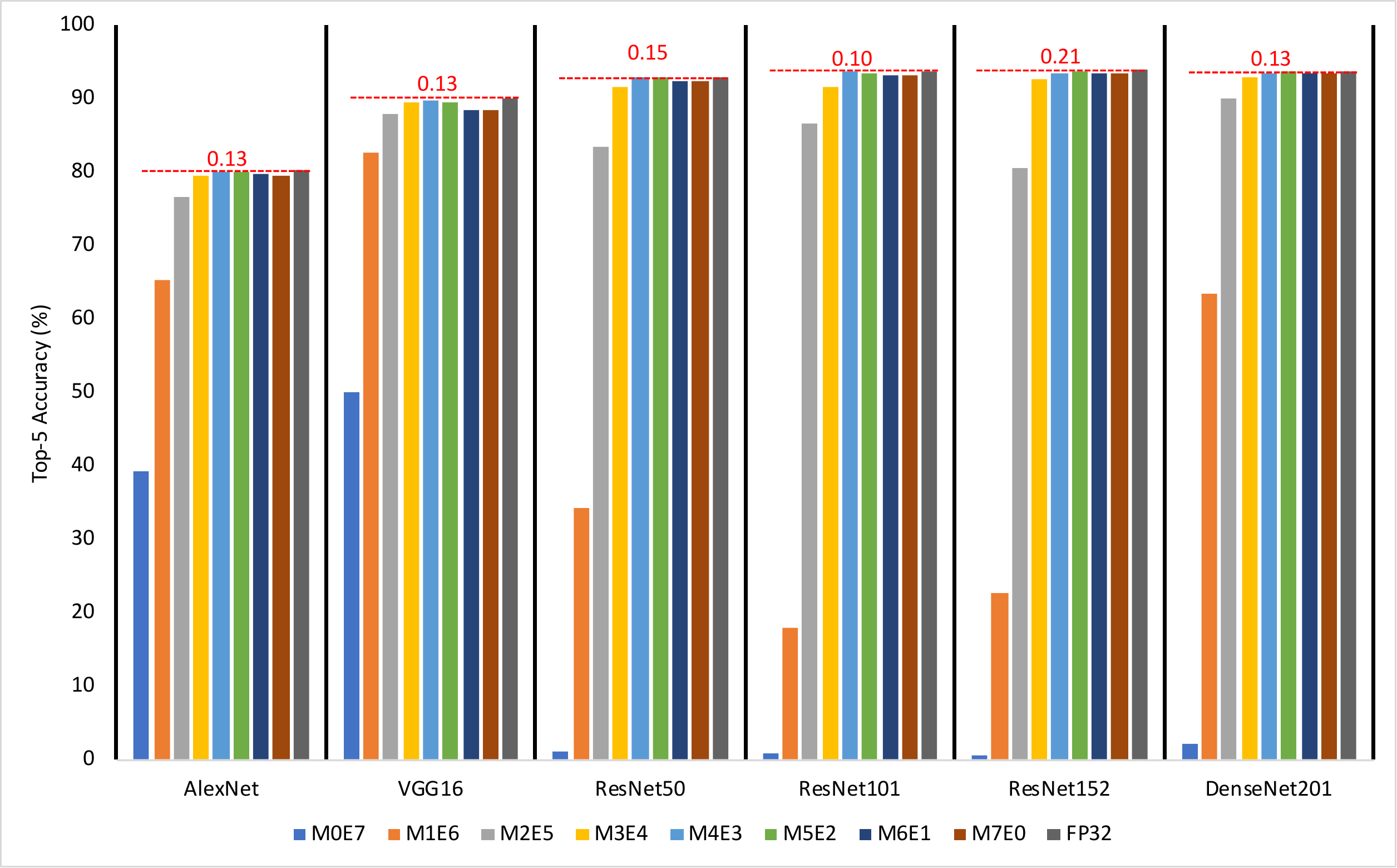}
\caption{Top-5 accuracy for different (mantissa, exponent) combinations with respect to different CNNs.}
\label{fig:top_5}
\end{figure}

Six representative CNNs including the {\it slim, medium} and {\it deep} CNNs are evaluated, as listed in Table~\ref{table:benchmark}. The detailed validation accuracy on the quantized network with all the benchmarks are shown in Figures~\ref{fig:top_1} and ~\ref{fig:top_5}. We emulate all 8 different (mantissa, exponent) combinations to validate accuracy of the quantized CNNs. The 32-bit floating-point results are included as the baseline.

\begin{table}[b]
  \centering
  \caption{Characteristics of CNN benchmarks. GOP is giga-operations needed by one 224$\times$224 RGB image.}
  \begin{tabular}{c|c|c|c}  \hline \hline
    {\bf CNN}              &{\bf Type}         &{\bf Operations}   &{\bf Model Weights}   \\ \hline
    AlexNet                &{\it slim}         &2.27 GOP           &249.51 MB             \\ \hline
    VGG16                  &{\it slim}         &30.94 GOP          &553.43 MB             \\ \hline
    ResNet50               &{\it medium}       &9.74  GOP          &46.05 MB              \\ \hline
    ResNet101              &{\it medium}       &19.70 GOP          &166.37 MB             \\ \hline
    ResNet152              &{\it deep}         &29.39 GOP          &229.39 MB             \\ \hline
    DenseNet201            &{\it deep}         &10.85 GOP          &68.63 MB              \\ \hline
  \end{tabular}
  \label{table:benchmark}
\end{table}

In Figures~\ref{fig:top_1} and ~\ref{fig:top_5}, the dashed lines illustrate the 32-bit floating-point baseline, while the values above the dashed lines are the accuracy loss compared with the baseline. One can see that our proposed approach can maintain top-1 and top-5 accuracy comparable to the baseline. On average, the top-1 and top-5 accuracy loss is within 0.5\% and 0.3\% compared with the full precision results. Particularly, $M5E2$ always achieves the highest accuracy compared with the other cases. For the cases with more than or equal to 3-bit mantissa, they maintain a low accuracy loss for all the six CNNs, while the case with less than 3-bit mantissa can hardly find accurate results. We also compare our proposed approach with the fixed-point situation, marked as $M7E0$ in the figures ($M7E0$ means 7-bit mantissa and no exponent, exactly fixed-point). As shown in Figures~\ref{fig:top_1} and ~\ref{fig:top_5}, $M4E3$ and $M5E2$ outperforms the fixed-point for all six benchmarks. This is consistent with our observation that non-uniform quantization fits better than fixed-point quantization. 

$M4E3$ and $M5E2$, which achieve the two best accuracies among all the test cases, are also compared with five typical approaches. We report both the top-1 and top-5 accuracy for all six benchmarks in Table~\ref{table:acc_cmp}, where "-" indicates no reported results in the literatures. We use the normalized top-1 accuracy in Table~\ref{table:acc_cmp} for the approaches proposed by ARM \cite{ARM_mixed} and Xilinx \cite{xilinx_float} as reported in their paper. The top-1 and top-5 accuracy in Table~\ref{table:acc_cmp} show that our 8-bit floating-point quantization method without any re-training can outperform the literatures in most cases. Moreover, besides the approach proposed by Nvidia \cite{nvidia_fix}, our method is the only one that can reach {\it deep} networks. 

On the other hand, $M4E3$ also has a comparable top-1 and top-5 accuracy compared with the 32-bit floating-point baseline. This is one reason why we design our processor with the $M4E3$ case. The other reason is that 8-bit floating-point multiplier implemented with $M4E3$ costs less area and gains higher working frequency than $M5E2$ does, which will be discussed in detail in Subsection~\ref{subsection:hardware}.

\begin{table*}
\centering
\caption{Accuracy comparison between $M4E3$, $M5E2$, references and FP32. "-" means no reported results. Results for ARM and Xilinx are converted actual accuracy from the normalized accuracy reported in their papers without circuit implementations, while others are based on circuit implementations.}
  \begin{tabular}{c|cc|cc|cc|cc|cc|cc}  \hline \hline
    \multirow{2}{*}{}             &\multicolumn{12}{c}{Top-1 Accuracy (\%) \quad Top-5 Accuracy (\%) for each network} \\ \cline{2-13}
                                  &\multicolumn{2}{c|}{AlexNet}   &\multicolumn{2}{c|}{VGG16}     &\multicolumn{2}{c|}{ResNet50} 
                                  &\multicolumn{2}{c|}{ResNet101} &\multicolumn{2}{c|}{ResNet152} &\multicolumn{2}{c}{DenseNet201}     \\    \hline
    Nvidia \cite{nvidia_fix}      &57.05      &80.06              &70.84      &-                  &73.10      &91.06
                                  &74.40      &91.73              &74.70      &91.78              &-          &-                       \\    \hline
    ARM \cite{ARM_mixed}          &56.71      &-                  &70.38      &-                  &-          &-
                                  &-          &-                  &-          &-                  &-          &-                       \\    \hline
    Xilinx \cite{xilinx_float}    &-          &-                  &-          &-                  &75.80      &-
                                  &-          &-                  &-          &-                  &-          &-                       \\    \hline
    BFP \cite{bfp_fpga}           &-          &-                  &68.32      &-                  &72.76      &-
                                  &-          &-                  &-          &-                  &-          &-                       \\    \hline
    V-Quant \cite{V-Quant}        &56.24      &78.95              &71.77      &90.66              &-          &- 
                                  &-          &-                  &78.35      &93.95              &77.32      &93.51                   \\    \hline
    FP32 (Baseline)               &57.28      &80.18              &70.38      &89.81              &75.80      &92.90
                                  &77.10      &93.70              &77.60      &93.83              &76.85      &93.62                   \\    \hline
    {\sf Phoenix ($M4E3$)}        &{\bf 56.69}&{\bf 79.99}        &{\bf 70.05}&{\bf 89.68}        &{\bf 75.25}&{\bf 92.75}
                                  &{\bf 76.68}&{\bf 93.60}        &{\bf 76.79}&{\bf 93.44}        &{\bf 76.40}&{\bf 93.43}             \\    \hline
    {\sf Phoenix ($M5E2$)}        &{\bf 56.77}&{\bf 80.05}        &{\bf 69.74}&{\bf 89.49}        &{\bf 75.37}&{\bf 92.71}
                                  &{\bf 76.43}&{\bf 93.33}        &{\bf 77.05}&{\bf 93.62}        &{\bf 76.55}&{\bf 93.49}             \\    \hline
  \end{tabular}
  \label{table:acc_cmp}
\end{table*}

\subsection{Normalization Analysis}
\label{subsection:normalization_analysis}

\begin{figure}[t]
\centering
\includegraphics[width=3.5in]{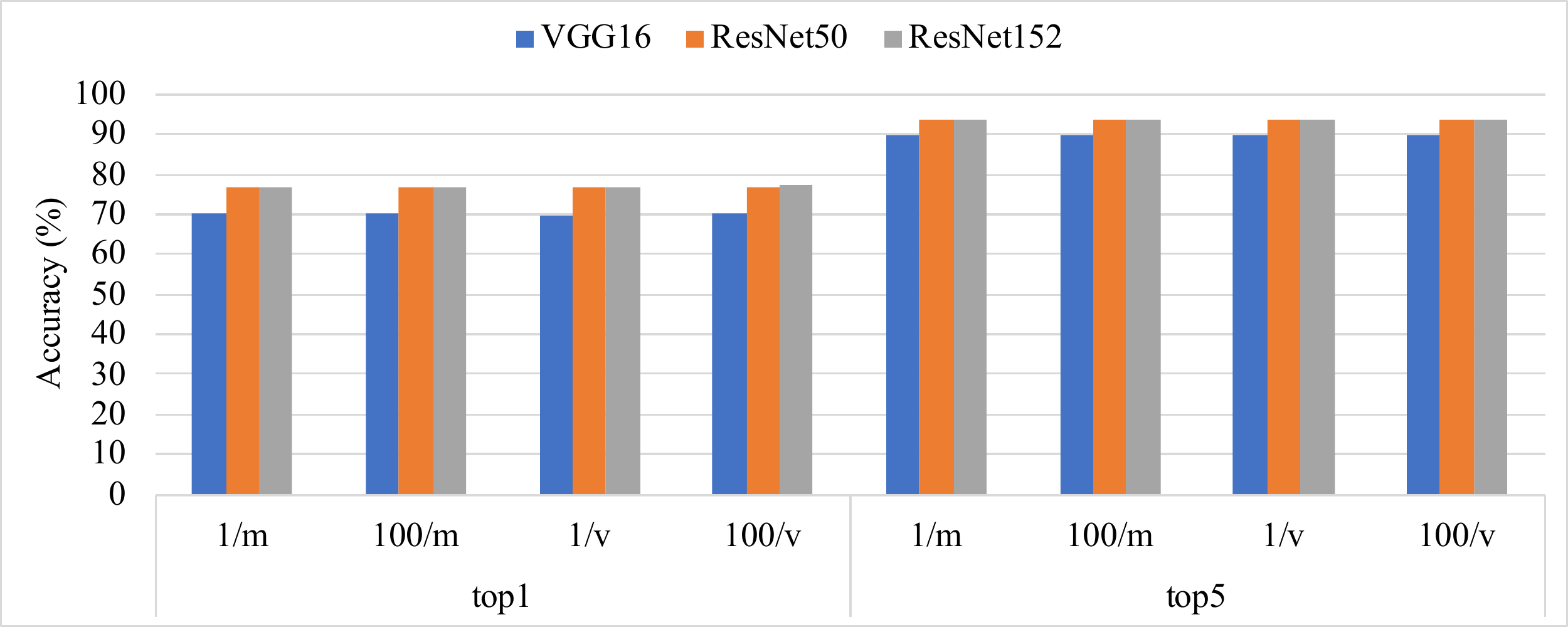}
\caption{Top-1 and Top-5 accuracy with different normalization methods and different number of test images (1/m: normalize with 2nd moment and 1 test image; 100/m: normalize with 2nd moment and 100 test images; 1/v: normalize with mean and standard deviation and 1 test image; 100/v: normalize with mean and standard deviation and 100 test images).}
\label{fig:norm_eva}
\end{figure}

In this subsection, different normalization methods are explored considering their influence on accuracy. During normalization, we first use a mini-batch of 100 test images to fetch the normalization parameter. This is because the normalization parameter is defined as the second moment of the output feature map, and more data can better describe its statistical characteristics. We then define the normalization parameter as the mean and standard deviation of the output feature map, and the normalization process is changed from Eq. ~(\ref{equation:normlization}) to Eq. ~(\ref{equation:norm_var}).
\begin{equation}
\label{equation:norm_var}
    NORM\_O_i^m = \frac{O_i^m - \mu(O^m)}{\sigma(O^m)},
\end{equation}
where $\mu(O^m)$ and $\sigma(O^m)$ denote the mean and standard deviation of the outputs of layer $m$. During the evaluation process, both one test image and a mini-batch of 100 test images are applied to this normalization method. We take three representative networks -- VGG16, ResNet50 and ResNet152 -- for {\it slim, medium} and {\it deep} networks as driving examples. The evaluation results are shown in Figure~\ref{fig:norm_eva}. One can see that normalization with a mini-batch of 100 test images outperforms that with one image for all cases. On average, the improvement of applying a mini-batch is 0.05\% for both top-1 and top-5 accuracy. However, the average quantization time of applying a mini-batch increases by 50$\times$, {\it e.g.,} from 17s to 983s for ResNet152. The normalization method with mean and standard deviation turns out to have almost the same accuracy results as the normalization method with second moment. In this way, we select the normalization method with second moment, since this incurs fewer efforts in merging all the normalization parameters. 

\section{Processor Architecture}
\label{section:processor_architecture}

In this section, we discuss in detail the architecture of {\sf Phoenix}, which efficiently overcomes the challenge of hardware inefficiency caused by floating-point based operations.

\subsection{Overview}
\label{subsection:overview}

\begin{figure}[t]
\centering
\includegraphics[width=3.5in]{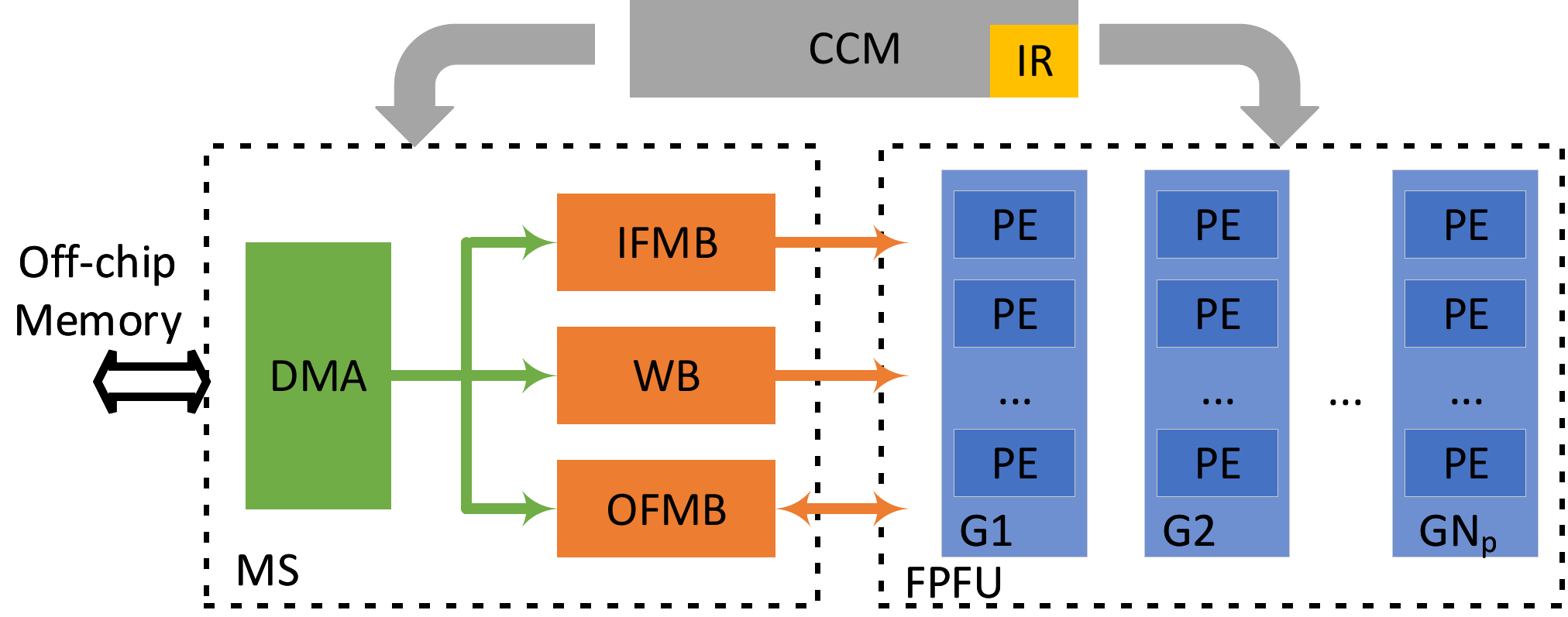}
\caption{The Architecture of {\sf Phoenix}.}
\label{fig:overview}
\end{figure}

The architecture of {\sf Phoenix} is depicted in Figure~\ref{fig:overview}. We develop a floating-point function unit (FPFU), which is composed of multiples of processing elements (PEs), to compute the outputs of a layer in parallel. The PE, which is the key component of {\sf Phoenix}, is designed to efficiently perform multiplications and additions of 8-bit floating-point data. The on-chip memory system (MS) consists of three buffers: input feature map buffer (IFMB), weight buffer (WB) and output feature map buffer (OFMB). All these three buffers are ping-pong architecture to hide the communication time between on-chip and off-chip memory through direct memory access (DMA) module. We design a central control module (CCM) to arbitrate between different modules. The CCM decodes various instructions stored in the instruction RAM (IR) into detailed signals for other modules. 

\subsection{Architecture of PE}
\label{subsection:PE}

\begin{figure}[t]
\centering
\includegraphics[width=3.5in]{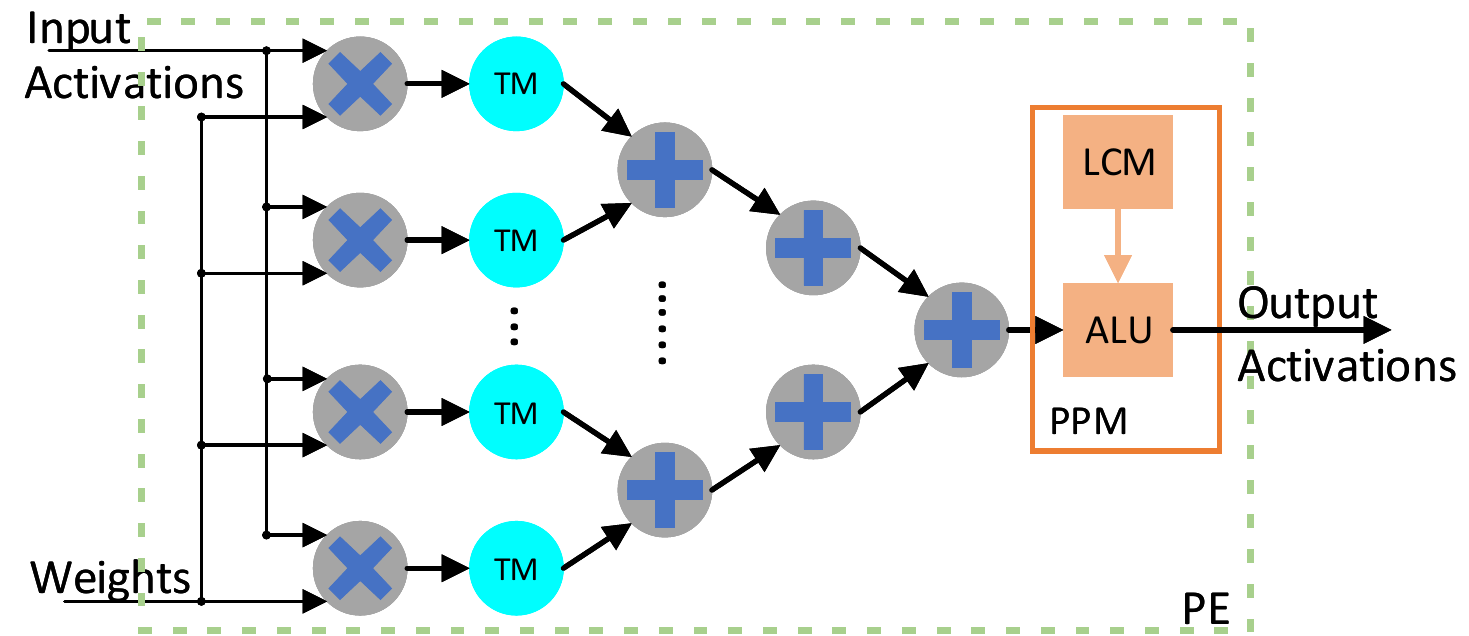}
\caption{The architecture of a PE.}
\label{fig:PE}
\end{figure}

{\sf Phoenix} is designed to perform floating-point multiplications and additions efficiently for performance gain and energy reduction. In {\sf Phoenix}, floating-point numbers are operated inside the PEs, which construct the FPFU. The FPFU receives activations from IFMB and weights from WB, then distributes to different PEs. Each PE operates the dot product of two vectors and stores the results into OFMB. Inside each PE, we design a fully pipelined data-flow-based architecture, as shown in Figure~\ref{fig:PE}. Once a PE receives two vectors, it distributes the data to $N_m$ multipliers, whose full-precision results are transferred into a truncating module (TM). The full-precision data are aligned to have same scale and truncated into low-precision to simplify the design of the adder tree, which is followed to sum up all the products. A post process module (PPM) then accumulates, activates and stores the data into the OFMB.

\subsubsection{8-bit floating-point multiplier}
8-bit floating-point numbers are represented with scientific notations in the sign-and-magnitude format, as illustrated in Eqs. ~(\ref{equation:8_float}) and ~(\ref{equation:subnormal}). The multiplication of two numbers is then divided into three fixed-point components: (1) XOR of the signs; (2) multiplication of mantissas; (3) addition of exponents. Take the $MaEb$ format as an example. An $(a+1)$-bit unsigned multiplier and a $b$-bit unsigned adder are designed inside each 8-bit floating-point multiplier of the PE. We design the multiplier to be $(a+1)$-bit because the first bit of mantissa is hidden for saving storage -- "1" for normal numbers and "0" for subnormal numbers. Meanwhile, $bias$ is not included during addition, because this is the same for all the numbers and we can address this at the last step to simplify the adders. The $2a+b+4$ bits result is arranged in the {\it sign-product-sum} order and delivered into the truncating module.

\subsubsection{Truncating module}
\begin{figure}[t]
\centering
\includegraphics[width=3.5in]{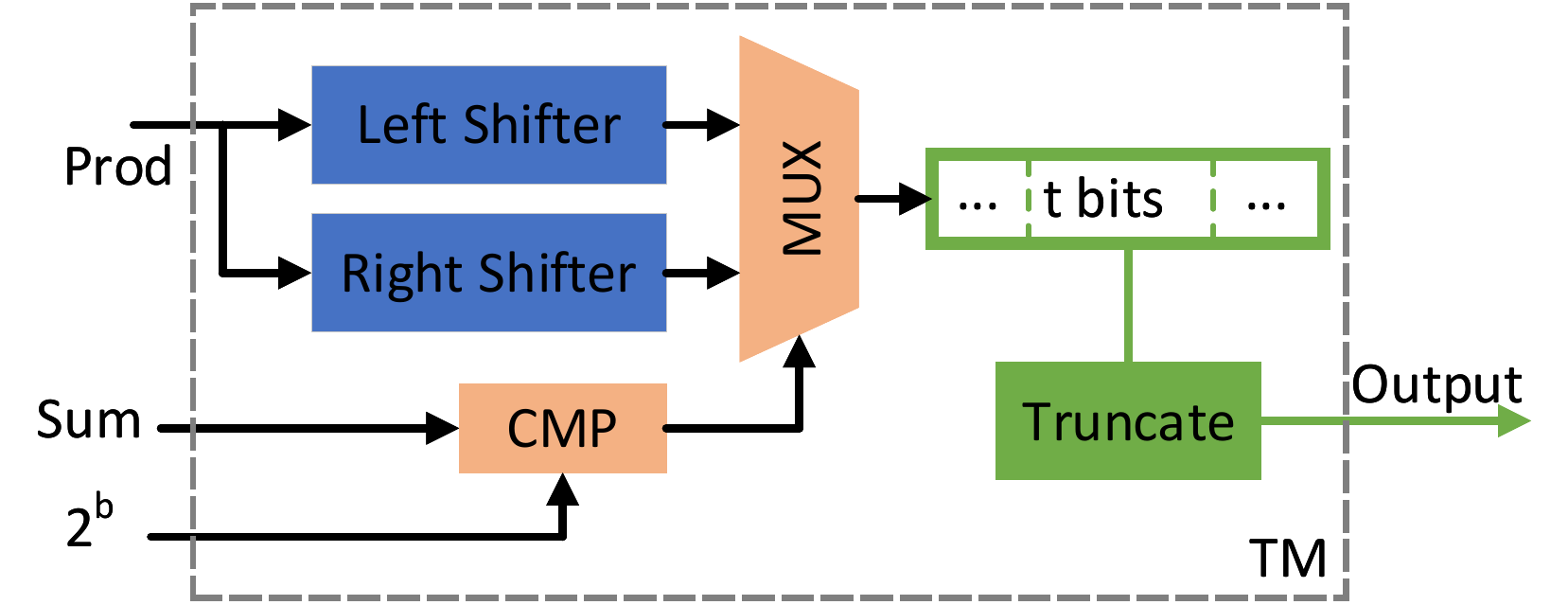}
\caption{Truncating process.}
\label{fig:TM}
\end{figure}

The full-precision results from the 8-bit floating-point multipliers incur high burden for adders because the results are in different scales, and only data in the same scale can be summed up. Therefore, we develop a truncating module to first align the data to the same scale and then truncate them into low-precision with $t$ bits to simplify the design of adders (the selection of $t$ will be explained in Subsection~\ref{subsection:hardware} with experimental results). In the alignment and truncation process, the dot position is kept unchanged, and the data width is truncated to $t$ bits according to the exponent. As shown in Figure~\ref{fig:TM}, we set the scale to $2^b$ and the {\it sum} will be compared with $2^b$ to decide whether the product is shifted left or right. This is the alignment process to keep the full precision. The aligned value is then cut into $t$ bits, where we consider saturation with overflow and round when discarding least significant bits. In this way, the adder tree can be implemented with fixed-point adders, which is more hardware efficient than floating-point adders. 

\subsubsection{Post-processing module}
The post-processing module consists of a local controller module (LCM) and an arithmetic logic unit (ALU). Controlled by the CCM of our processor, LCM indicates the control signals for computing a layer, as well as the activation parameters. The ALU is designed to perform simple calculations, {\it e.g.,} add biases, activate with ReLU, accumulate intermediate results and convert the output to 8-bit floating-point format. 

\subsection{Memory System}
\label{subsection:ms}

The memory system in our processor is divided into three parts: an input feature map buffer (IFMB), an output feature map buffer (OFMB) and a weights buffer (WB), as shown in Figure~\ref{fig:overview}. All the buffers are designed in ping-pong manner to hide the time of DMA memory accesses under the time of computation.

The width of IFMB is set to be $N_g \times N_m \times 8$ bits as to provide $N_m$ 8-bit floating-point input activations to $N_g$ PEs each time. The $N_g$ PEs are grouped into one PE group to focus on $N_g$ output activations on the same output feature map, which share the same weights. In our processor, we design $N_p$ PE groups, where the input activations are shared. In this way, the width of WB is chosen to be $N_p \times N_m \times 8$ bits to provide weights for $N_p$ different output channels. The OFMB is then set to have the width of $N_p \times N_g \times 16$ bits to save $N_p \times N_g$ output activations. Although each pixel in the output feature map is represented with 8-bit floating-point, we keep the intermediate results with 16-bit precision to reduce accuracy loss.

The parameters $N_g, N_m$ and $N_p$ are decided to trade off between area, overall performance and energy, which are discussed in Subsection~\ref{subsection:hardware}. The sizes of the three buffers are also decisive to the area, overall performance and energy. Previous proposed work applies large enough buffers to store all the activations or weights for one layer \cite{EIE} to avoid costly off-chip memory access. However, such designs incur large area and unscalability for larger and deeper CNNs. In our processor, we trade off among the area, scalability, performance and energy, and employ the smallest sizes which can hide the DMA communication time. After exploring different CNNs and buffer sizes, we deploy 64KB, 64KB and 32KB for IFMB, OFMB and WB, respectively. During inference on our processor, only when all the input feature map have been processed and reused, or all the weights have been processed and reused, or OFMB is full, will the off-chip memory be accessed for loading new input feature maps, loading new weights or storing output feature maps, respectively.

\subsection{Central Control}
\label{subsection:central_ctrl}

The CCM is designed to arbitrate among different modules and control the whole execution process. Firstly, it decodes the instructions from IR efficiently and sets the corresponding control registers. Then, different modules are activated according to the control registers and the status of each module are monitored by the control registers as well. Finally, the CCM decides when to fetch the next instruction from the feedback of the control registers. We also design a compiler to generate the block-level instructions.

\section{Experimental Methodology}
\label{section:experimental methodology}

In this section, we introduce the experimental methodology. We develop {\sf Phoenix} in Verilog and then synthesize, place and route it with IC Compiler using TSMC 28nm library. The energy cost is evaluated with the PrimeTime PX tool based on the waveform files obtained from post-implementation simulation. The off-chip memory access energy is estimated by using the tools provided by MICRON \cite{dram}. We also design a cycle accurate simulator to estimate the throughput for different CNNs. 

{\bf Baselines}. We select the CPU, the GPU and customized accelerators as baselines.

{\bf CPU and GPU}. We use the Darknet framework to evaluate the benchmarks on an Intel (R) Core (TM) i9-7960X CPU working under 2.86GHz. We also use darknet to evaluate the benchmarks on a Nvidia TITAN Xp GPU, which has a 12GB DDR5. Furthermore, we use the cuBLAS \cite{cuBLAS} to implement the benchmarks on the GPU.

{\bf Customized Accelerator}. We compare {\sf Phoenix} against three customized accelerators: 8-bit fixed-point processor, Eyeriss \cite{eyeriss} and OLAccel \cite{OLAccel}. The 8-bit fixed-point processor is implemented in the same architecture and with the same parameters as that of {\sf Phoenix}. We select Eyeriss as another baseline because it provides an open-source estimation tool for comparison \cite{eyeriss_jssc}. We perform an ISO-area comparison between Eyeriss and {\sf Phoenix}. Thus, we scale {\sf Phoenix} from TSMC 28nm to TSMC 65nm, which is the technology node used by Eyeriss. We allocate the same core area and the same amount of on-chip memory. We first get the core area of 168 8-bit PEs in Eyeriss. Then the number of PEs for our processor is decided to meet the area target, and the configuration is shown in Table~\ref{table:config}. OLAccel does not provide any open-source estimation tool, but reports their comparison results with Eyeriss. Therefore, we compare our results with the reported results in OLAccel.

\begin{table}[b]
  \centering
  \caption{Configurations of Eyeriss and {\sf Phoenix}.}
  \begin{tabular}{ccc}                                                               \hline
                        &{\bf Eyeriss}             &{\bf {\sf Phoenix}}           \\ \hline \hline
    \# of PEs           &168                       &768                           \\
    core area ($mm^2$)  &0.96                      &1.03                          \\ \hline
    On-chip Memory      &181.5KB                   &WB:51.5KB                     \\
                        &                          &IFMB, OFMB: 64KB              \\ \hline
  \end{tabular}
  \label{table:config}
\end{table}

{\bf Benchmarks}. The six CNNs listed in Table~\ref{table:benchmark} are used as benchmarks for comparison with the CPU and the GPU. Only {\it convolutional} layers in AlexNet and VGG-16 are utilized for comparison with Eyeriss, since Eyeriss only supports {\it convolutional} layers. 

\section{Experimental Results}
\label{section:experimental results}

\begin{figure}[t]
\centering
\includegraphics[width=3.5in]{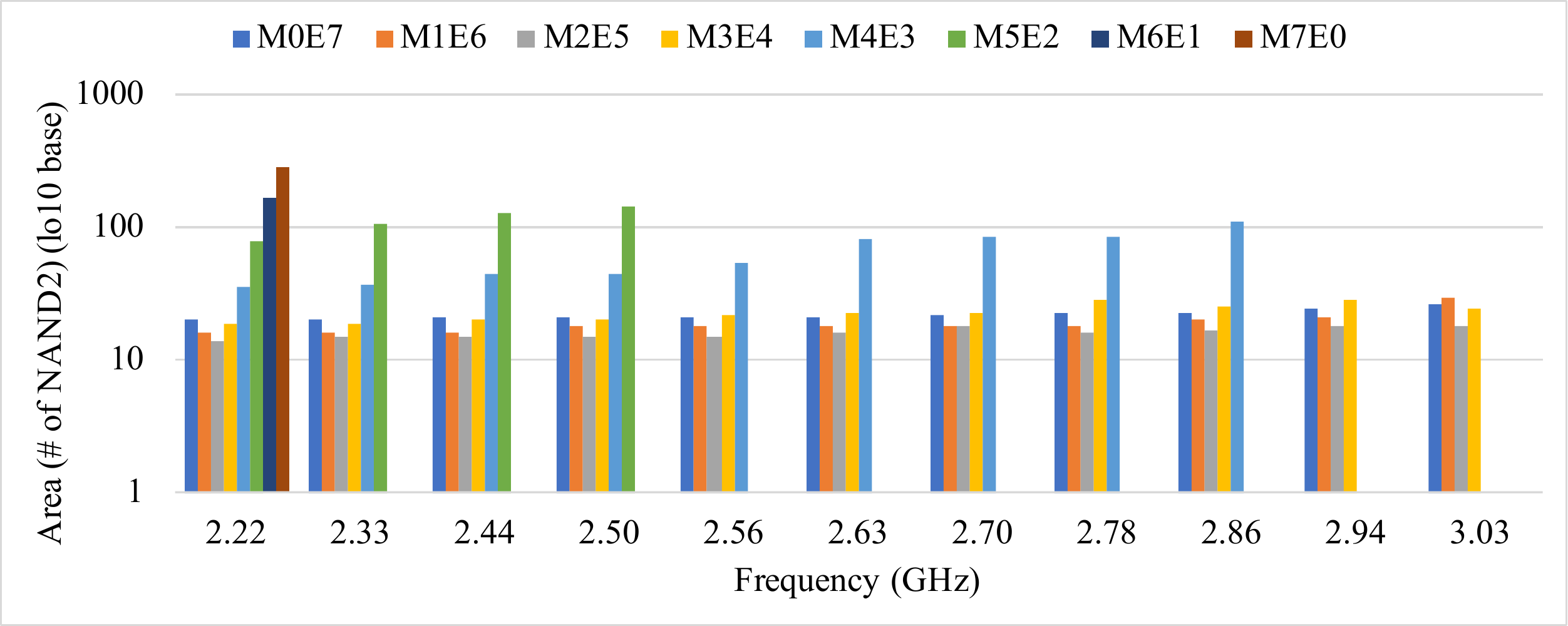}
\caption{Area (equivalent to the number of NAND2) for 8-bit floating-point multiplier with respect to frequency.}
\label{fig:mul}
\end{figure}

\subsection{Hardware Characteristics}
\label{subsection:hardware}

\subsubsection{Multiplier}
\label{subsubsection:multiplier}
We explore all the eight (mantissa, exponent) combinations for 8-bit floating-point multipliers before the implementation of our processor. We implement multipliers with Verilog and synthesize them with Synopsys design compiler to evaluate the maximal working frequency and corresponding area at TSMC 28nm technology node, as shown in Figure~\ref{fig:mul}. $M4E3$ and $M5E2$ are able to work under 2.86GHz and 2.5GHz with the area equivalent to 112 and 146 2-input NAND gates, respectively. The maximal frequency for $M6E1$ and $M7E0$ (also known as fixed-point version) is 2.22GHz, while the area is equivalent to 169 and 285 2-input NAND gates, respectively. As for the cases with less than 4-bit mantissa, they all reach the frequency of 3.03GHz, while the corresponding area is 26, 29, 18 and 24 2-input NAND gates, respectively. As discussed in Subsection~\ref{subsection:quantization_results}, $M4E3$ and $M5E2$ outperform all the other cases for accuracy. In terms of working frequency and area, $M4E3$ is better than $M5E2$. Particularly, the maximal frequency of $M4E3$ is 14.4\% higher than that of $M5E2$. As to the area, $M4E3$ is 3.24$\times$ more efficient than $M5E2$ when working under $M5E2$'s maximal frequency. Meanwhile, compared with the fixed-point (marked as $M7E0$ in the figure) version, $M4E3$ reduces the area by 8.14$\times$ at the same working frequency. Although the case $M3E4$ outperforms $M4E3$ in both working frequency and area, the top-1 and top-5 accuracy loss of $M3E4$ are 3.73$\times$ and 3.48$\times$ higher than that of $M4E3$, respectively. This is not acceptable since maintaining accuracy is one of the key motivations in our work. 

\begin{figure}[t]
\centering
\includegraphics[width=3.5in]{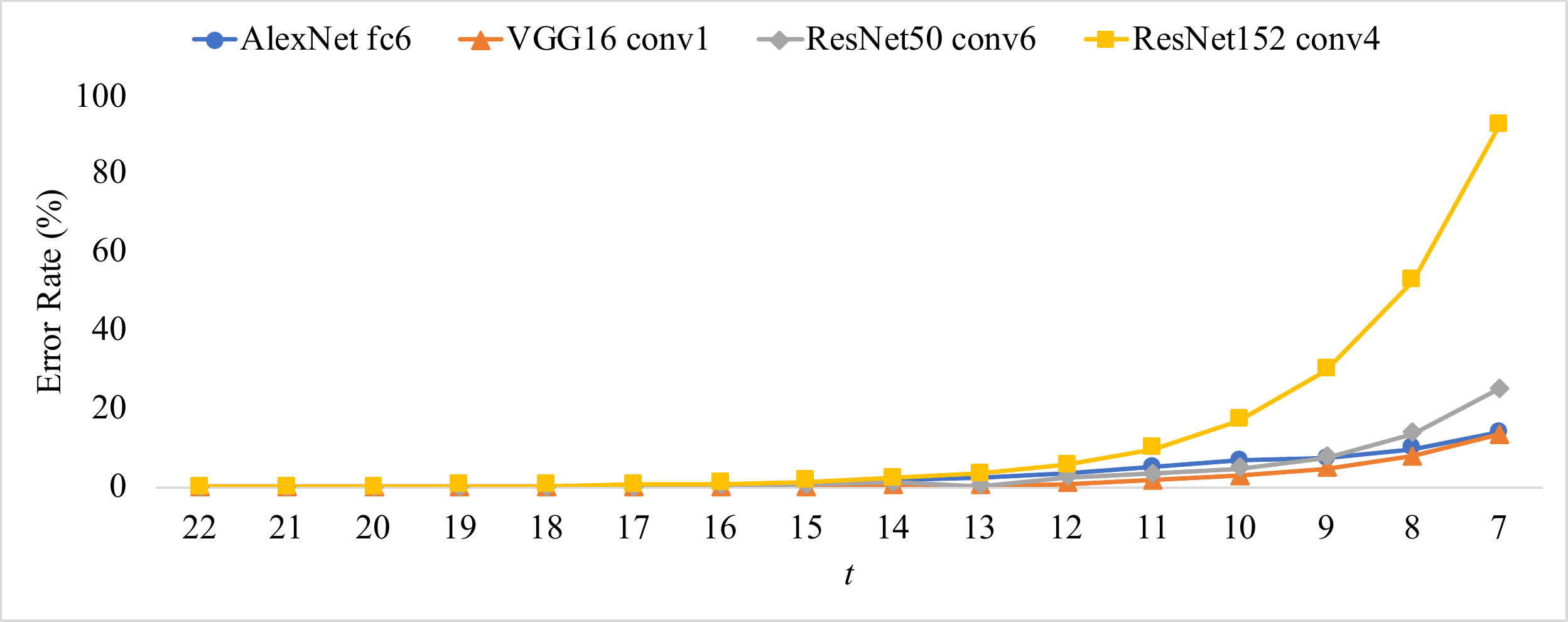}
\caption{Error rate compared with the quantized network for different layers with respect to different truncating parameter $t$.}
\label{fig:err_rate}
\end{figure}

\begin{figure}[t]
\centering
\includegraphics[width=3.5in]{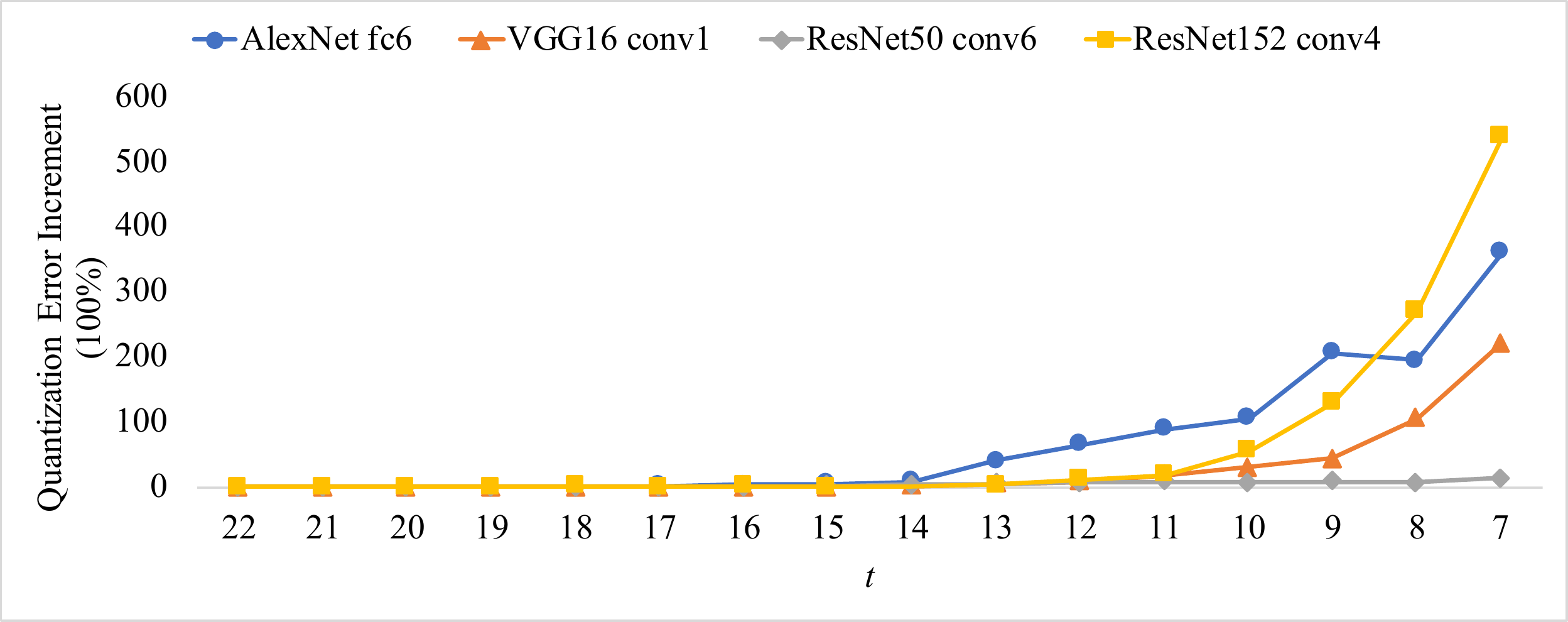}
\caption{Error rate increment with respect to different truncating parameter $t$.}
\label{fig:err_incre}
\end{figure}

\subsubsection{Truncating Bit-width}
\label{subsubsection:truncating_bit}
In the truncating module (TM), we cut the full precision number into $t$ bits to simplify the adder design. To convert the product of two $M4E3$ numbers to the same scale in full precision, we need at least 22 bits. This is because the mantissa of the product is 10 bits, and the exponent ranges from 2 to 14. Note that we exploit different $t$s to evaluate the errors incurred by reducing precision. We take four representative layers -- {\it fc6} in AlexNet, {\it conv1} in VGG16, {\it conv6} in ResNet50 and {\it conv4} in ResNet152 -- as driving examples. We depict the error rate for output activations of each layer caused by reducing $t$ from 22 to 7, as shown in Figure~\ref{fig:err_rate}. Compared with quantized networks, reducing $t$ from 22 to 14 incurs little error to activations of the layers (less than 1\%). However, when $t$ is less than 14, the error increases dramatically, e.g., the error increases from 4.38\% to 92.68\% for {\it conv4} in ResNet152, which is not acceptable. This advantage comes from our quantization method. During the quantization process, we normalize all activations and use a Gaussian distribution to approximate the activations. When we truncate it from 22 bits to 14 bits, we discard large values located outside the $3\sigma$ region, which has less impact on the accuracy of the layer than that of the values within the $3\sigma$ region. Moreover, when $t$ is less than 14, more values inside the $3\sigma$ region are discarded, which results in a higher error rate.

We further explore the influence on quantization error when introducing the truncating technique, as shown in Figure~\ref{fig:err_incre}. When $t$ changes from 22 to 14, this error remains in the same level. However, when $t$ decreases from 14 to 7, the quantization error increases significantly, e.g., the quantization error increases by 3.57\% to 538.30\% for {\it conv4} in ResNet152. This is not acceptable because it will incur large accuracy loss. Two experimental results demonstrate that if we truncate the bit width of activations from 22 bits to 14 bits, we will not suffer from large quantization error. Therefore, we select $t=14$ in our current design. We speculate that in general $t$ can be selected based on $3\sigma$ of Gaussian distribution in this study.

\subsubsection{Memory System Parameters}
\label{subsubsection:ms_param}
\begin{figure}[t]
\centering
\includegraphics[width=3.5in]{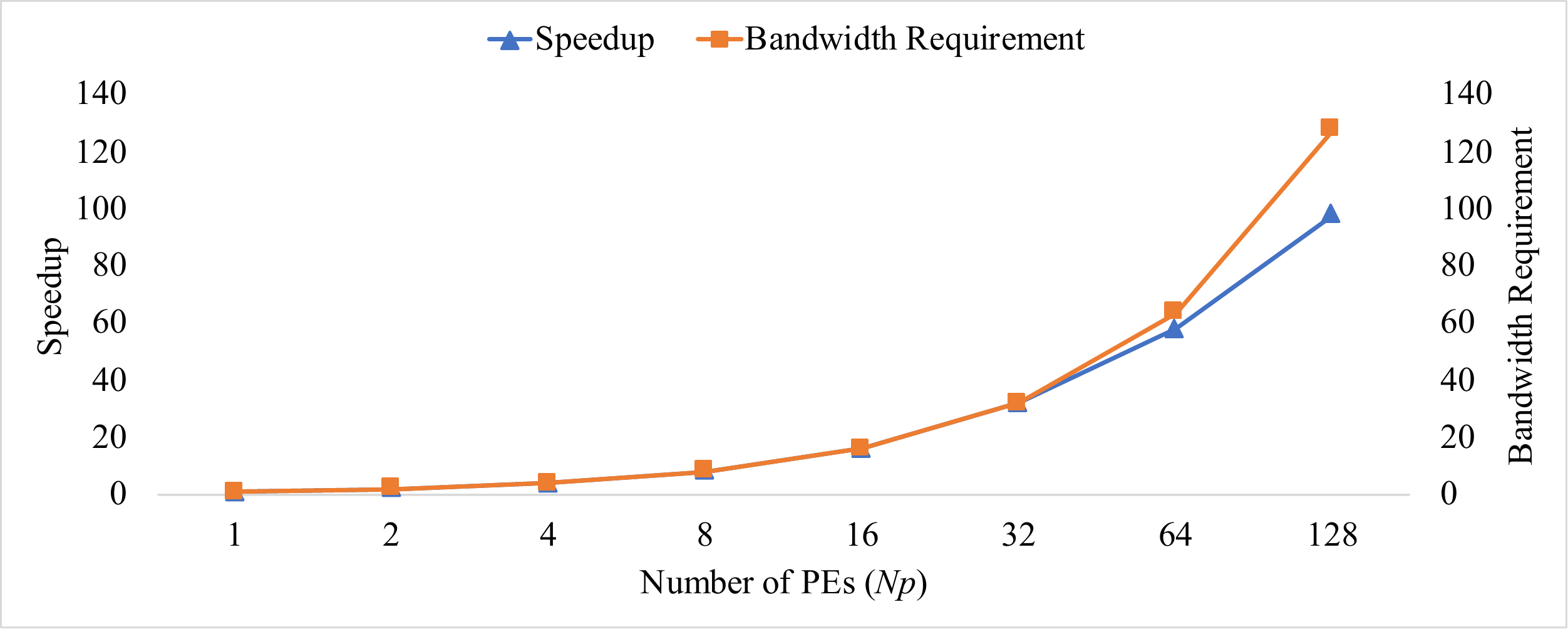}
\caption{Speedup and bandwidth requirement versus number of PEs.}
\label{fig:scale}
\end{figure}

In our current design, we select $N_g = 4$ which is also efficient with small feature map size. In CNNs based on the ImageNet data set, the smallest size of a feature map is always $14\times14$ or $8\times8$, which indicates that calculating 4 pixels in parallel achieves the highest efficiency. The parameter $N_m$ is chosen to be 32 with the consideration of resource utilization. Most CNNs have their channel number to be multiples of 32, except the first layer. In the first layer, the channel number is 3 as they have RGB images as inputs. When $N_m > 32$, e.g, $N_m = 64$, 50\% of the multipliers will be wasted for the layers like {\it conv1}, {\it conv2} in AlexNet, {\it conv1} in VGG-16, as the channel number of these layers is 32 or 96. At the same time, $N_p$ is decided by considering the resource utilization and the memory bandwidth. As shown in Figure~\ref{fig:scale}, we explore the speedup and bandwidth requirement for VGG-16 when enlarging $N_p$ from 1 to 128 by multiplying 2 each time. When the $N_p$ is less than 64, the speedup doubles as $N_p$ doubles. This is because all PEs are working efficiently. However, the speedup turns out to have a saturation when $N_p >= 64$. This is because large amount of PEs are wasted when the output channel number is less than 64. Furthermore, the bandwidth requirement doubles when $N_p$ doubles as the on-chip memory need to provide enough data for all PEs. However, the increment in bandwidth also enlarges the buffer size to hide the time of off-chip memory access, which will incur larger area and energy. Therefore, we select $N_p=16$ in our current design to trade-off between performance, resource utilization, area and energy. 

\subsubsection{Implementation}
\label{subsubsection:implementation}
{\sf Phoenix} is implemented in TSMC 28nm technology node, and the results after placement and routing are listed in Table~\ref{table:impl}. At 0.9V, the peak throughput is 2.048 TMAC/s (TMACS) with a 1GHz core clock rate. The core area is $1.44 mm^2$ with the total power of $1091.2 mW$. Among them, the memory system (including IFMB, WB, OFMB) consumes 50.7\% of the total power. 

\begin{table}[b]
  \centering
  \caption{The results of {\sf Phoenix} after placement and routing.}
  \begin{tabular}{|c|c|}                                                          \hline
    {\bf Category}                                &{\bf Parameters}           \\  \hline \hline
    Technology                                    &TSMC 28nm HPC+ 1P10M       \\  \hline
    Core Size                                     &$1.2\times1.2 mm^2$        \\  \hline
    Core Power                                    &$1091.2 mW$                \\  \hline
    \# of MACs ($N_m \times N_g \times N_p$)      &2048 \\  \hline
    Supply Voltage                                &Core 0.9V                  \\  \hline
    Clock Rate                                    &1GHz                       \\  \hline
    Peak Throughput                               &2.048TMAC/s                \\  \hline
    Arithmetic Precision                          &8-bit floating-point       \\  \hline
  \end{tabular}
  \label{table:impl}
\end{table}

\subsection{Performance}
\label{subsection:performance}
We compare the execution time of {\sf Phoenix} against CPU and GPU on the six CNNs listed in Table~\ref{table:benchmark}. On CPU and GPU, we evaluate the CNNs with both 8-bit floating-point and 32-bit floating-point precision (i.e., CPU-8, GPU-8, CPU-32, GPU-32).

\begin{figure}[t]
\centering
\includegraphics[width=3.5in]{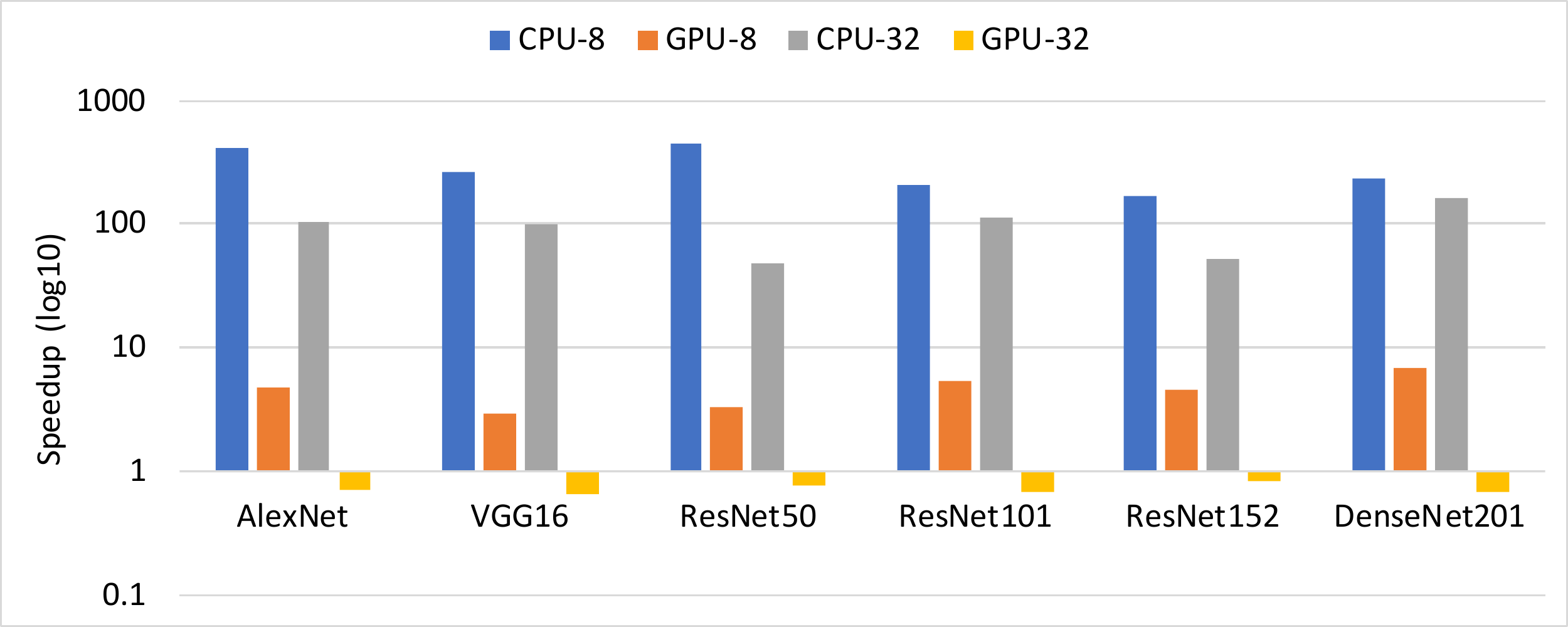}
\caption{The speedup of {\sf Phoenix} over Intel (R) Core (TM) i9-7960X CPU and Nvidia TITAN Xp GPU.}
\label{fig:speedup}
\end{figure}

We normalize the execution time against that of {\sf Phoenix} to gain the speedup, as shown in Figure~\ref{fig:speedup}. Compared with CPU and GPU in 8-bit precision, {\sf Phoenix} achieves 290.7$\times$ and 4.7$\times$ speedup, respectively. This is because CPU and GPU do not have any optimization on 8-bit floating-point operations. During the evaluation, we make the conversions between 8-bit floating-point and 32-bit floating-point representations for each layer on both CPU and GPU, which incurs a large time burden. We also evaluate the 32-bit precision on CPU and GPU, and {\sf Phoenix} is still 95.6$\times$ faster than CPU and is 70\% as fast as Titan Xp. Such speedups against CPU mainly come from the FPFU modules in {\sf Phoenix}, where we optimize multiplications and additions of the 8-bit floating-point number. Although {\sf Phoenix} achieves 70\% of the speed of Titan Xp, it consumes less energy than Titan Xp does, as explained in Subsection~\ref{subsection:energy}.

\subsection{Energy}
\label{subsection:energy}
\begin{figure}[t]
\centering
\includegraphics[width=3.5in]{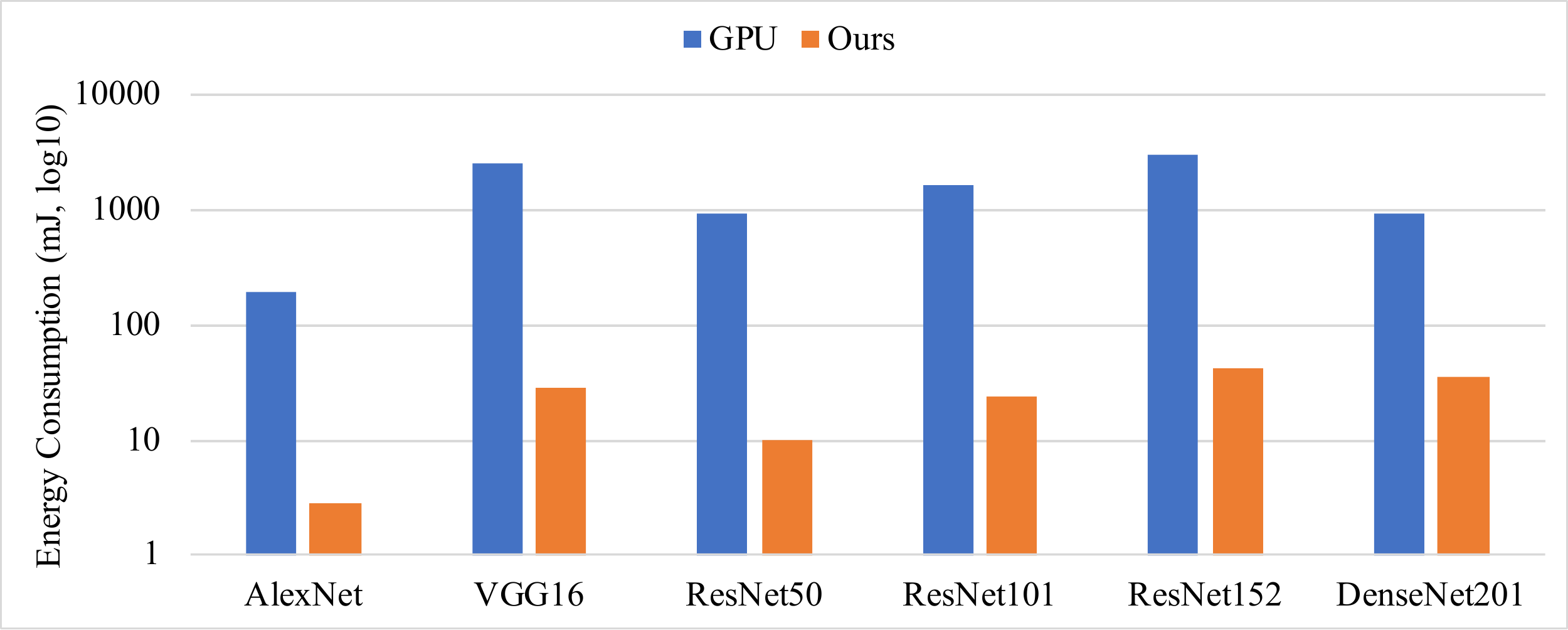}
\caption{Energy comparison between Nvidia TITAN Xp GPU and {\sf Phoenix}.}
\label{fig:energy}
\end{figure}

We report the energy comparison between {\sf Phoenix} and Titan Xp across all the six benchmarks, as shown in Figure~\ref{fig:energy}. We include the off-chip memory access energy in this comparison. Compared with GPU, {\sf Phoenix} achieves 70.5$\times$ better energy efficiency on average. This advantage comes from the 8-bit floating-point quantization, which reduces the memory amount in 4$\times$, thus reducing the number of memory accesses. Moreover, {\sf Phoenix} is designed to reuse both the weights and activations, which also leads to the reduction in off-chip memory accesses. Regarding the processor energy cost without off-chip memory access, we can achieve 151$\times$ compared with GPU. This results demonstrate the high energy efficiency of {\sf Phoenix}.

\begin{figure}[t]
\centering
\includegraphics[width=3.5in]{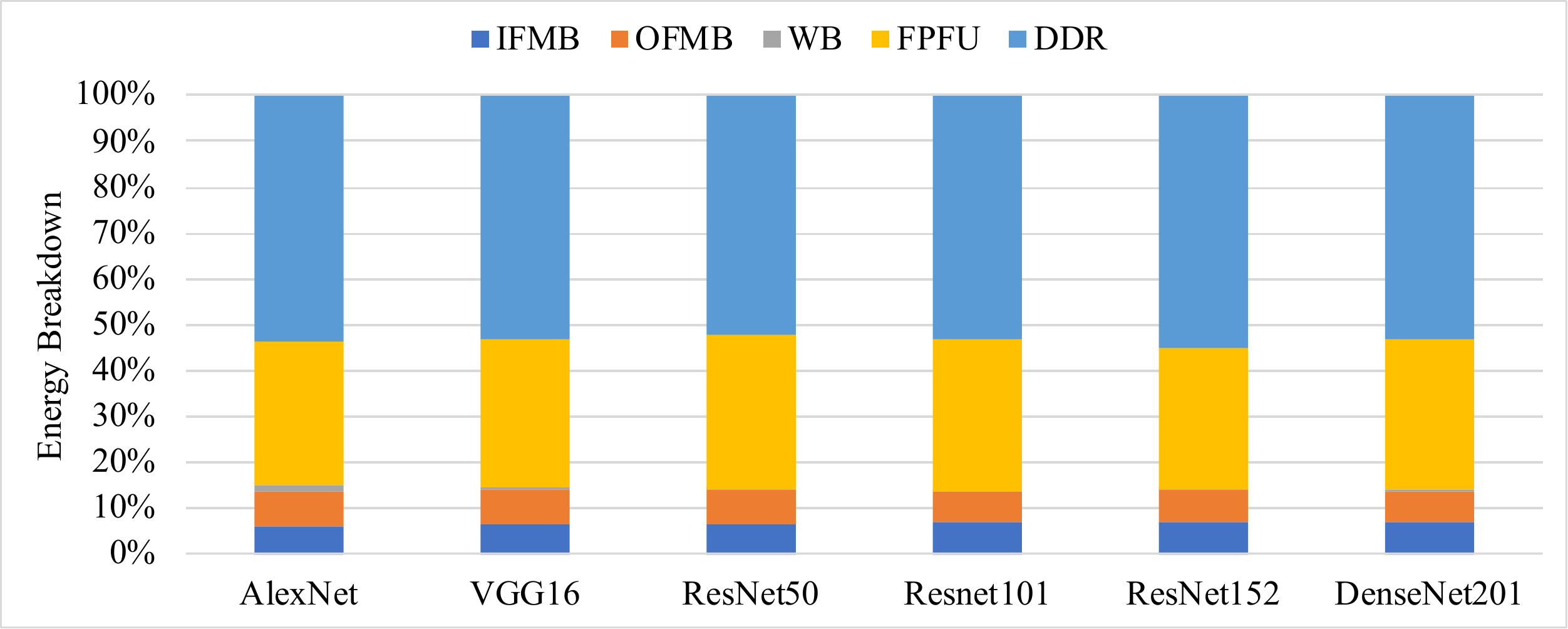}
\caption{Energy breakdown of {\sf Phoenix} with off-chip memory accesses.}
\label{fig:breakdown_ddr}
\end{figure}

Meanwhile, the energy breakdown of {\sf Phoenix} with off-chip memory access for all evaluated benchmarks is shown in Figure~\ref{fig:breakdown_ddr}. We can observe that the energy consumed by off-chip memory access is more than 50\%. This is because we still need to transfer large amount of data from the off-chip memory to the on-chip memory system. The result also informs us that quantization and data reuse are the key points for reducing the energy caused by off-chip memory access.

\begin{figure}[t]
\centering
\includegraphics[width=3.5in]{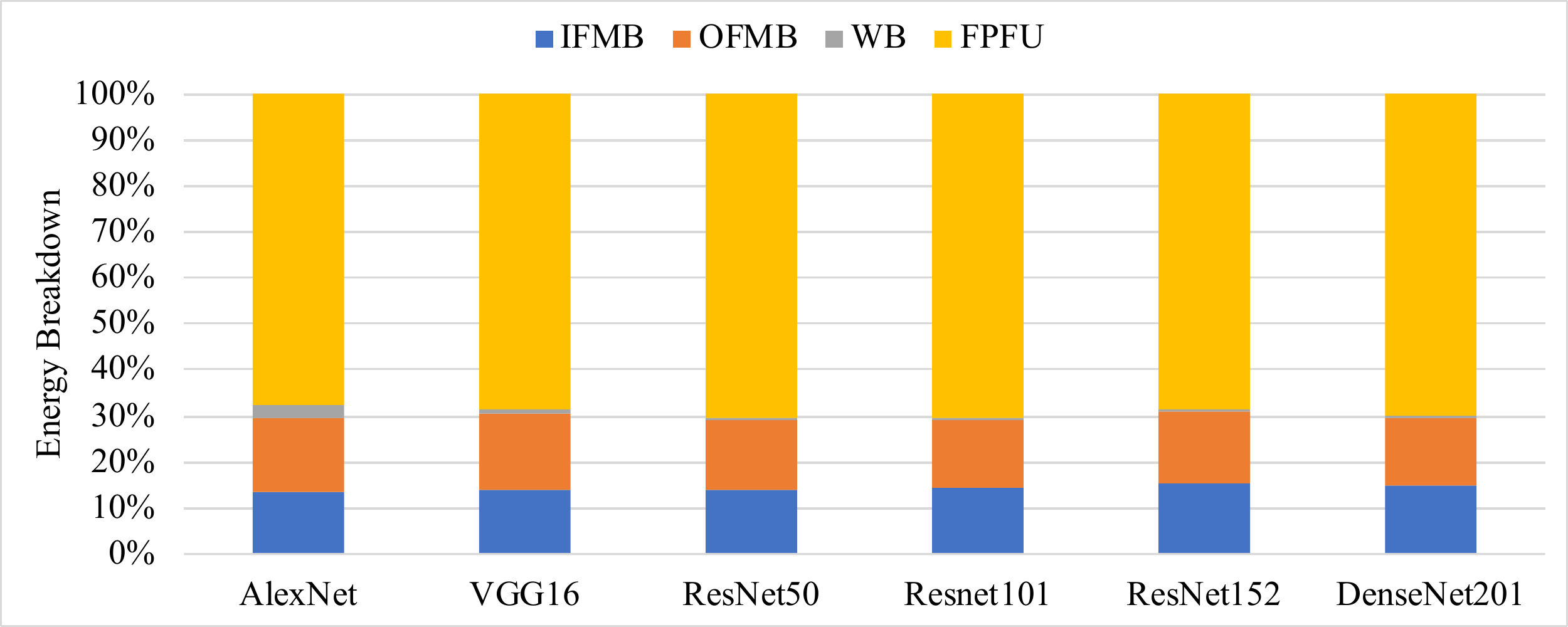}
\caption{Energy breakdown of {\sf Phoenix} without off-chip memory accesses.}
\label{fig:breakdown}
\end{figure}

We further present the energy breakdown without off-chip memory access in Figure~\ref{fig:breakdown}. The on-chip memory system (including IFMB, WB and OFMB) consumes about 30\% of the total energy. The WB consumes the least energy, because we reuse the weights for all the input activations stored in IFMB, which reduces the accesses for WB. Moreover, the WB consumes more energy in AlexNet and VGG-16 than that of other benchmarks. This is because these two networks have relatively larger {\it fully-connected} layers, which benefits less in {\sf Phoenix}. As for the FPFU, it consumes about 70\% of the total energy. The FPFU, which is a fully-pipelined data-flow-based architecture, works all the time when inputs exist. We do not skip zero activations nor weights in our design, which makes the FPFU consume more energy.

\subsection{Discussion}
\label{subsection:discussion}

\subsubsection{Comparison with Fixed Point Processor}
We also compare {\sf Phoenix} with the 8-bit fixed-point processor with the same architecture and the same parameters. Hence, the two processors have the same performance. Based on this, {\sf Phoenix} uses 1.2$\times$ less area than the 8-bit fixed-point processor according to the DC synthesize results, and both processors have similar energy. Although the fixed-point processor does not need the truncating module and also benefits from the data reuse technique, the 8-bit fixed-point multipliers consumes more area than 8-bit floating-point multipliers as explained in Subsubsection~\ref{subsubsection:multiplier}.

\subsubsection{Comparison with Other Accelerators}
{\sf Phoenix} is also compared with a state-of-the-art fixed-point accelerator, Eyeriss, as it provides the open-sourced estimation tool \cite{eyeriss_jssc}. For fair comparison, we scale {\sf Phoenix} to have the same area and the same on-chip memory as Eyeriss in 65nm technology node, as shown in Table~\ref{table:config}. We only compare the computation time for {\it convolutional} layers, because Eyeriss only supports {\it convolutional} layers. All {\it convolutional} layers in AlexNet and VGG-16 are compared in Tables~\ref{table:Eye_alex} and ~\ref{table:Eye_vgg}. 

\begin{table}[b]
  \centering
  \caption{Performance comparison against Eyeriss for AlexNet ($ms$).}
  \begin{tabular}{|c|c|c|c|}  \hline
    {\bf Layer}     &{\bf Eyeriss}      &{\bf {\sf Phoenix}}  &{\bf Speedup}       \\ \hline \hline
    CONV1           &4.1                &0.8                  &5.13$\times$        \\ \hline
    CONV2           &9.8                &3.2                  &3.06$\times$        \\ \hline
    CONV3           &5.5                &1.1                  &5.00$\times$        \\ \hline
    CONV4           &4.0                &1.6                  &2.50$\times$        \\ \hline
    CONV5           &2.5                &1.1                  &2.27$\times$        \\ \hline
    {\bf Total}     &{\bf 25.9}         &{\bf 7.8}            &{\bf 3.32$\times$}  \\ \hline
  \end{tabular}
  \label{table:Eye_alex}
\end{table}

One can seen from Tables~\ref{table:Eye_alex} and ~\ref{table:Eye_vgg}, {\sf Phoenix} has a speedup of 3.32$\times$ and 7.45$\times$ for AlexNet and VGG-16, respectively. This is mainly because we have more multipliers for the same area as Eyeriss (768 in {\sf Phoenix} versus 168 in Eyeriss). In {\sf Phoenix}, we flatten input RGB images according to the kernel size to save computation time. Therefore, we have larger speedup for the first {\it convolutional} layer for both benchmarks. As for layers with small input feature map size, e.g. CONV5 in AlexNet ($13\times13$) and CONV5-1, CONV5-2, CONV5-3 in VGG-16 ($14\times14$), the advantages in performance reduce. In these layers, the whole input feature map can be stored in the IFMB, hence {\sf Phoenix} benefits less from data reuse than the layers with large input feature map size. 

We further compare with the OLAccel\cite{OLAccel}, whose hardware is based on the V-Quant quantization method. It does not provide any open-sourced estimation tools so that we compare with their reported performance comparison results with Eyeriss. Regarding AlexNet and VGG-16, OLAccel outperforms Eyeriss with 3.55$\times$ and 5.56$\times$ in performance, respectively. Compared with OLAccel, our processor does not have better performance in AlexNet, because they skip zero activations during calculation, which saves execution cycles. However, this does not work on larger CNN. Our processor outperforms OLAccel in VGG-16 with a 1.34$\times$ reduction in execution cycles. There are two reasons for that: 1) {\sf Phoenix} has more multiplier units than OLAccel (768 in {\sf Phoenix} versus 576 in OLAccel); 2) The quantization method used in OLAccel requires keeping around 3\% of the weights and activations with full-precision, which brings the burden to the hardware and the execution process. 

\begin{table}[t]
  \centering
  \caption{Performance comparison against Eyeriss for VGG-16 ($ms$).}
  \begin{tabular}{|c|c|c|c|}  \hline
    {\bf Layer}     &{\bf Eyeriss}      &{\bf {\sf Phoenix}}  &{\bf Speedup}      \\ \hline \hline
    CONV1-1         &12.7               &0.8                  &15.88$\times$      \\ \hline
    CONV1-2         &270.2              &23.7                 &11.40$\times$      \\ \hline
    CONV2-1         &135.1              &13.2                 &10.23$\times$      \\ \hline
    CONV2-2         &270.3              &26.5                 &10.20$\times$      \\ \hline
    CONV3-1         &68.0               &11.3                 &6.02$\times$       \\ \hline
    CONV3-2         &136.0              &22.7                 &5.99$\times$       \\ \hline
    CONV3-3         &136.0              &22.7                 &5.99$\times$       \\ \hline
    CONV4-1         &35.0               &7.3                  &4.79$\times$       \\ \hline
    CONV4-2         &70.0               &14.5                 &4.83$\times$       \\ \hline
    CONV4-3         &70.0               &14.5                 &4.83$\times$       \\ \hline
    CONV5-1         &16.1               &3.6                  &4.47$\times$       \\ \hline       
    CONV5-2         &16.2               &3.6                  &4.50$\times$       \\ \hline
    CONV5-3         &16.2               &3.6                  &4.50$\times$       \\ \hline
    {\bf Total}     &{\bf 1251.8}       &{\bf 168}            &{\bf 7.45$\times$} \\ \hline
  \end{tabular}
  \label{table:Eye_vgg}
\end{table}

\section{Related Work}
\label{section:related work}
CNNs are typically over-parameterized, and extensive studies in recent years focus on CNN approximation algorithms, including weight reduction and quantization \cite{approx_survey}. A deep compression method was proposed in \cite{han_pruning, han_compression}. Weight pruning along with iterative re-training was first applied to CNNs to reduce weights \cite{han_pruning}, after which weights were quantized using k-means clustering. The quantized network was then re-trained to compensate for quantization error. Finally, they used Huffman coding to represent the quantized weights to save memory \cite{han_compression}.

A lot of researchers concentrate on quantization to save storage. Binarization quantizes parameters into just two values, typically \{-1, 1\} with a scaling factor \cite{bin1, bin2, bin3, xnor_net}. Although binarization achieves remarkable energy and storage saving, it suffers from significant accuracy loss when binarizing both weights and activations \cite{bin1, bin2, bin3}. Among them, XNOR\_Net \cite{xnor_net} can maintain comparable accuracy for AlexNet by only applying weights binarization. Ternary representations, which adds zero to the binary set, were introduced to help improve the accuracy \cite{ter1, ter2}. Logarithmic based quantization was proposed in \cite{log_quan}. Their results showed that 5-bit weights with log-base $\sqrt2$ and 5-bit activations with log-base 2 can achieve 2.72\% top-1 accuracy loss and 1.71\% top-5 accuracy loss for VGG-16. Quantization with 8-bit fixed-point is one general way to maintain low accuracy loss \cite{google_fix, qual_fix}. The authors explored the relationship between the bit width and accuracy by quantizing the weights with floating-point and activations with fixed-point \cite{ARM_mixed}. Intel \cite{intel_quan} also tried to find out the relationship between the bit width and accuracy. Their experiments also included one floating-point quantization case along with other fixed-point cases. In \cite{xilinx_float}, the authors explored dynamic floating-point based quantization with a calibration process to compensate for the accuracy loss in their work. However, all the aforementioned work failed to show promising results for \textit{deep} CNNs such as ResNet152.

The approach in \cite{nvidia_fix} showed that 8-bit fixed-point quantization was possible for \textit{deep} CNNs. In \cite{entropy_quan}, the authors quantized the weights with 5-bit and activations with 6-bit using their weighted-entropy-based quantization method for ResNet101, which achieved a small accuracy loss. A more aggressive method \cite{V-Quant} provided promising results for \textit{deep} CNNs. They quantized the small values of the weights into 4 bits while remained the rest 16 bits as full precision, by dividing the weights into the low-precision and high-precision regions according to the values of the weights. However, these work all need extra components to address the full precision weights. Different from all the above methods, our quantization methods fully exploit the distribution of weights and activations, thus obtaining a comparable or better accuracy for \textit{deep} CNNs without any extra calibration, fine-tuning or re-training that need labelled data and extra computing. Access to labelled data could be difficult in practice as hardware and CNN algorithms are often developed by different parties.

CNN accelerators benefit a lot from the approximation algorithms with respect to improving energy efficiency and throughput. EIE \cite{EIE}, Cambricon-X \cite{cambricon_x} and Cambricon-S \cite{cambricon_s} are the ones that use weights reduction techniques. They build sparse matrix oriented architectures to accelerate CNNs after deep compression. However, they need re-training to compensate for accuracy loss. In addition, they need extra hardware components to address the irregularity caused by sparsity. FINN \cite{FINN} and FP-BNN \cite{FP_BNN} are two binarization based accelerators. Although they can achieve higher energy efficiency and throughput than the full-precision counterparts, they both have a high accuracy loss. The accelerator in \cite{log_quan} was developed to speedup the quantized CNN with log-based weights and activations. Stripes \cite{Stripes} and Bit Fusion \cite{bit_fusion} performed layer-wise mixed-precision inference using bit-serial MACs. However, they failed to exploit \textit{deep} CNNs. OLAccel was developed based on the V-Quant method \cite{OLAccel}, which used a large amount of 4-bit MACs plus a small portion of mixed-precision MAC units to cope with the high-precision region. This leads to significant energy saving compared with 8-bit fixed-point accelerators. However, their quantization method also need re-training to compensate for the quantization error. In addition, they keep a small portion of full-precision weights and activations during their quantization process, which leads to a hardware overhead to cope with the full-precision values. Overall, \textsf{Phoenix} is better in terms of performance and energy efficiency.

\section{Conclusion}
\label{section:conclusions}
In this paper, we propose a normalization-oriented 8-bit floating-point quantization method which saves memory storage and memory access as well as maintaining negligible accuracy degradation (less than 0.5\% for top-1 and 0.3\% for top-5 accuracy). Due to the use of normalization, our quantization method gets rid of extra cost for calibration, fine-tuning or re-training to compensate for accuracy loss. We further design a hardware processor named {\sf Phoenix} to fully leverage the benefits of our proposed quantization method and address the hardware inefficiency caused by floating-point operations. The key feature of {\sf Phoenix} is the fully pipelined data-flow-based PE, which shares input activations and weights with other PEs, thus reducing inner bandwidth requirements. The circuit placement and routing results show that {\sf Phoenix} can achieve peak performance of 2.048TMAC/s with 1.44$mm^2$ and 1091.2$mW$ at TSMC 28nm technology, respectively. Compared with a state-of-the-art accelerator, {\sf Phoenix} achieves 3.32$\times$ and 7.45$\times$ better performance with the same core area for AlexNet and VGG-16, respectively. Compared with Nvidia TITAN Xp GPU, {\sf Phoenix} consumes 151$\times$ less energy with single image inference.

% or


\ifCLASSOPTIONcaptionsoff
  \newpage
\fi



\bibliographystyle{IEEEtran}
\bibliography{refs}
\end{document}